\documentclass[12pt]{article}
\usepackage{amsmath,amssymb}
\usepackage{bm}
\usepackage{graphicx}
\usepackage{setspace}
\usepackage{natbib}
\usepackage{nameref,hyperref}
\usepackage{authblk}

\usepackage{courier}
\usepackage{pdflscape}
\usepackage{url}
\usepackage{amsmath}
\usepackage{soul}
\usepackage[usenames,table]{xcolor}
\usepackage{gensymb}
\usepackage{lineno}

\usepackage{multirow}
\newcommand{\phix}{$\phi \chi$174}
\newcommand{\ddg}{$\Delta \Delta G$}
\newcommand{\micro}{$\mu$}
\newcommand{\milli}{m}

\definecolor{light-gray}{gray}{0.95}

\begin{document}

\title{Evolutionary interplay between structure, energy and epistasis in the coat protein of the $\phi  \chi$174 phage family}

\author[1]{R.A.F.  Redondo \thanks{Both authors contributed equally to this work}} 

\author[1,2]{H.P. de Vladar,\thanks{Corresponding author. E-mail: Harold.Vladar@parmenides-foundation.org}$^*$}

\author[3]{T. W\l odarski}

\author[1]{J.P. Bollback}

\affil[1]{IST Austria. Am Campus 1. A-3400 Klosterneuburg, Austria.}
\affil[2]{Center for the Conceptual Foundations of Science, Parmenides Foundation. 82049 Pullach, Germany.}
\affil[3]{Department of Structural and Molecular Biology, University College London. WC1E 6BT London, United Kingdom.}

%\small{
%\noindent $^{1}$IST Austria. Am Campus 1. A-3400 Klosterneuburg, Austria.\\
%$^{2}$Center for the Conceptual Foundations of Science, Parmenides Foundation. 82049 Pullach, Germany.\\
%$^{3}$Department of Structural and Molecular Biology, University College London. WC1E 6BT London, United Kingdom.\\ 
%$\dag$ These authors contributed equally to this work.\\
%* Corresponding Author. E-mail: Harold.Vladar@parmenides-foundation.org
%}

\date{August 2015}

\maketitle

\newpage

\begin{abstract}
Viral capsids are structurally constrained by interactions amongst the amino acids of the constituting proteins. Therefore,  epistasis is expected to evolve amongst sites engaged in physical interactions, and to influence their substitution rates. In order to study the distribution of structural epistasis, we modeled \emph{in silico} the capsid of 18 species of the \phix \, family, including the wild type. \phix \, is amongst the simplest organisms, making it suitable for experimental evolution and \emph{in silico} modeling. We found nearly 40 variable amino acid sites in the main capsid protein across the 18 species. To study how epistasis evolved in this group, we reconstructed the ancestral sequences using a Bayesian phylogenetic framework. The ancestral states include 8 variable amino acids, for a total of 256 possible haplotypes. The $dN/dS$ ratio is low, suggesting strong purifying selection, consistent with the idea that the structure is constrained by some form of stabilizing selection. For each haplotype in the ancestral node and for the extant species we estimated \emph{in silico} the distribution of free energies and of epistasis. We found that free energy has not significantly increased but epistasis has. We decomposed epistasis up to fifth order and found that high-order epistasis can sometimes compensate pairwise interactions, often making the free energy seem additive. By synthesizing some of the ancestral haplotypes of the capsid gene, we measured their fitness experimentally, and found that the predicted deviations in the coat protein free energy do not significantly affect fitness, which is consistent with the stabilizing selection hypothesis.
\end{abstract}


\section{{Introduction}\label{sec:Intro}}

One of the central questions of evolutionary biology is to understand how the extant diversity of a group or species derives from their common ancestor through mechanisms such as selection, mutation, demographic events like population bottlenecks and others. 
Phylogenetics is a fundamental framework to study evolution based on, for example, molecular data of extant species. However, the specific biological reasons behind the diversification patterns that are observed and/or inferred by phylogenetics often remain elusive and, at times, unknown. This, in part, is due to the complexity of organisms and the uncertainty of  what selection is acting on, even in cases when molecular signatures strongly indicate the action of selection. Through phylogenetic analyses, signatures are sometimes found at the maximum resolution, i.e., at the nucleotide level. However, it is not clear that these nucleotides can be direct targets of selection. Selection can act on a complex genotype-phenotype map, favoring not specific alleles at specific loci, but complete traits encoded by multiple loci in a non-additive way, that is, when there are epistatic effects on a trait.

The aim of this work is to understand the evolution of the capsid of the bacteriophage \phix \, family. The phageÕs capsid is made up by several copies only four proteins: the coat and scaffold proteins and the major and minor spike proteins, which are direct product of translation (Fig. \ref{Fig:CapsidModel}). The capsid is structurally conserved across species; amongst the family of Microviridae, the molecular variation in the coat protein is low ($\lesssim$5\%), specifically when compared to the spike proteins that show substantial variation. The spike proteins can be so divergent that even alignments fail (even though, structure is conserved). In this work we focus specifically on the coat protein, which is the central structural constituent of the capsid.

By performing phylogenetic reconstructions of the coat protein of different members of this family we find that selection has been acting during the diversification of the phageÕs capsid. Despite the clear signatures of selection, the causing factors driving this evolution are not obvious. We therefore approach the problem also from the point of view of structural biology. We study the evolution of the free energies of the coat protein of different species of the \phix \, family. We estimate the phylogeny of this group and reconstruct the ancestral states for each amino acid (AA) at every internal node. From this data we determine which haplotypes to model and experimentally synthesize and assay their fitness effects, in order to study the evolution of the capsid.

By viewing the free energy of the capsid as an evolvable trait, we will ask how interactions amongst AAs result in epistatic effects. Although free energy is fully determined by the physical basis of the molecular structure and its environment, it is sequence dependent. Therefore, variability in sequences in the population of viruses result in a distribution of free energy values. Note that this distribution is not the distribution of molecular micro-states as described by statistical mechanics. For a given sequence there is a canonical distribution of energy states resulting on a particular value of free energy. Thus, on a fixed physical environment, a genetically heterogeneous population of viruses will have a distribution of free energies that is determined fully by the distribution of haplotypes. If the genetic composition of the population changes due to mutational and selective processes, so does the distribution of free energy, even though the latter is entirely determined physically. In other words, the folding free energy change that we study results from AA substitutions, not from structural processes such as folding.

We take advantage of both powerful frameworks: we use physical simulations to determine the free energy of a haplotype, and use phylogenetics to make inferences of evolutionary patterns of the haplotypes. From these analyses we will address the interplay between epistasis, selection and molecular evolutionary rates.

Despite the conspicuous signatures of selection that we observe in the gene of the coat protein, variation on the capsid structure is minimal, suggesting strong purifying selection. What causes this pattern? Compensatory mutations can be additive, but also non-additive, or epistatic in nature. Which of these are prevailing? Does the nature of the substitutions matter? These are the questions we address in this paper.

A closer look into the capsid reveals a certain amount of phenotypic complexity on the free energy of the different haplotypes. We estimate free energies \emph{in silico} using the know structure of the capsid of the wild type (WT) of the \phix \, phage (GeneBank accession J02482). The energy spectrum reveals the presence of non-additive effects, which we interpret as epistasis. This result is important because it is known that epistasis plays a significant role in masking structural and energetic variation \citep{Hermisson:2003dw,Lunzer:2010co,Breen:2012fd}. Free energy is understood as the capacity to do work. In our context, we estimate the free energy of structure unfolding for any  given haplotype. However, we do not consider major structural rearrangements. Instead, we keep the backbone of the proteins fixed and evaluate the free energy differences resulting from the substitution of AA, by allowing conformational changes only of the side chains. The free energy of the capsid is estimated based on physical principles that are independent from any assumption regarding evolution. These structural analyses provide independent information from that obtained from the phylogenetic analyses. By considering both, the phylogenetic and the structural information we can study how evolution maintains the structure of the capsid.

The next aspect that we address is how physical interactions amongst AAs determine the strength of epistasis. This gives insight on the distribution of non-additive effects on a trait, a standing question in evolution. We study high order epistasis, which is important because most works consider mostly pairwise interactions \citep{Weinreich:2013iy}. In general our knowledge of high order epistasis is limited. From the theoretical side, analyses of epistasis are largely based on pairwise interactions due to mathematical simplicity; high order epistasis results in complex models involving high-order terms and are thus intractable. Similarly, experimental studies often consider only pairwise interactions amongst genes due to methodological reasons, which include lack of statistical power coming from limited sample sizes \citep{Otwinowski:2014hk}. Consequently, we largely ignore how higher order epistasis can affect evolution. We address this point in this article. From simulated data we are able to go beyond pairwise effects and consider up to 5th order epistatic interactions. Importantly, we find that sometimes, high order interactions can compensate pairwise effects. This compensation can be deceiving, because some trait values that seem additive, can actually be highly epistatic. This result is critical because it reveals that we might often be neglecting important factors limiting evolution, such as mechanisms that potentially mask variation.
To understand the components of selection further, we also approach the subject experimentally. We synthesized the coat protein gene F of some of the ancestral haplotypes, and cloned them into the wild-type \phix \, genome replacing the WT version. We then measured the fitness of these individuals by infecting \emph{Escherichia coli} hosts. Surprisingly, we found no significant fitness differences relative to the WT phage.

We argue that together, the molecular analysis, biophysical calculations and experimental essays, provide evidence of stabilizing selection acting on the capsid of the \phix -like phages.

Although at first sight it seems that  free energy of unfolding is unrelated to evolutionary processes, we find convincing evidence of the action of stabilizing selection. This connection between structural biology and evolution provides direct evidence of structure-function relationships.

Our approach is suggestive of a larger scope regarding the role of epistasis in molecular evolution. Although it is intuitively obvious that the material basis on which selection acts can impose evolutionary constraints, it remains unclear how important its role can be.

\begin{landscape}
\begin{table}[H]
\setlength{\tabcolsep}{4pt}
\renewcommand{\arraystretch}{0.6}
\rowcolors{2}{light-gray}{white}
\begin{tabular}{cccccccccccc}
\hline \hline
Node & \multicolumn{8}{c}{Polymorphisms}  & \parbox[c]{1.5cm}{Nr.\\haplo.}&  \parbox[c]{2cm}{Most likely\\haplotypes} & Pr. \\ \hline
A	& K83Q	& T92S	& P141A	& E150Q	& Q153E	& Q182L	& S339A	& A361V	& 256	& \texttt{KSPEEQAA}		& 0.067	\\
B	&$\bigstar$&$\bigstar$&$\bigstar$&			&			&			&$\bigstar$&			& 16	& \texttt{KSAeeqSa} 		& 0.81	\\
C	& 	 		&			&			&			&			&			&			&			& 1(2)*	& \texttt{qsaeelaa} 		& 0.97* 	\\
D	&$\bigstar$&$\bigstar$&			&$\bigstar$&			&			&$\bigstar$&			& 16	& \texttt{KTpEeqAa} 		& 0.43 	\\
E	& 	 		&$\bigstar$/N/A$\dag$&	&		&			&			&			&			& 4		& \texttt{kNpqeqaa} 		& 0.46 	\\
F	& 	 		&$\bigstar$&			&$\bigstar$&			&			&$\bigstar$&			& 8 	& \texttt{kTpEeqSa} 		& 0.73$\ddag$ 	\\
G	& 	 		&$\bigstar$&$\bigstar$&$\bigstar$&$\bigstar$&			&			&$\bigstar$& 32	& \texttt{kTPEEqsA} 		& 0.68$\ddag$	\\
H	& 	 		&			&			&			&$\bigstar$&			&			&			& 2		& \texttt{ktpeEqsa} 		& 0.89$\ddag$ 	\\
I	& 	 		&			&			&$\bigstar$&$\bigstar$&			&$\bigstar$&			& 8(16)*	& \texttt{ktpEEqAa} 		& 0.424* 	\\
J	& 	 		&$\bigstar$&			&			&			&			&			&$\bigstar$& 4		& \texttt{kTaeeqsA} 		& 0.76	\\
K	& 	 		&			&			&			&			&			&			&			& 1		& \texttt{ktpeeqsa} 		& 0.99$\ddag$	\\
L	& 	 		&			&			&$\bigstar$&$\bigstar$&			&			&			& 4(8)*		& \texttt{ktpEEqaa} 		& 0.64*	\\
M	& 	 		&$\bigstar$&			&			&			&			&			&			& 2		& \texttt{kTpeeqsa} 		& 0.98$\ddag$	\\
N	& 	 		&			&			&$\bigstar$&			&			&			&			& 2(4)*	& \texttt{ktpEqqaa} 		& 0.51*	\\
O	& 	 		&			&			&			&			&			&			&			& 1		& \texttt{ktpeeqsv} 		& 0.99	\\
P	& 	 		&			&			&			&			&			&			&			& 1 	& \texttt{ktpqqqaa} 		& 0.99	\\
\hline \hline
\end{tabular}
\caption{Polymorphisms found in the ancestral reconstructions of each internal node of the Phylogenetic tree (Figure 1). Node A is the most recent common ancestor of the ingroup \phix \ species analyzed and presents the eight ancestral polymorphisms discussed in the text, the AWT is \textbf{ktpeqqsa}. The number of haplotypes and the most likely haplotype for each node with the posterior probability associated is also presented. Capital letters represents polymorphic sites and lower letters fixed sites in that node. * represent nodes with one additional polymorphism (G101R) not present in the ingroup ancestral Node A. $\dag$ Node E showed multiple alleles on site T92S/N/A. $\ddag$ Nodes whose most likely haplotype is identical to the consensus of the extant species.}
\label{Table:Polymorphisms}
\end{table}
\end{landscape}

\begin{figure}[h]
\includegraphics[width=\textwidth]{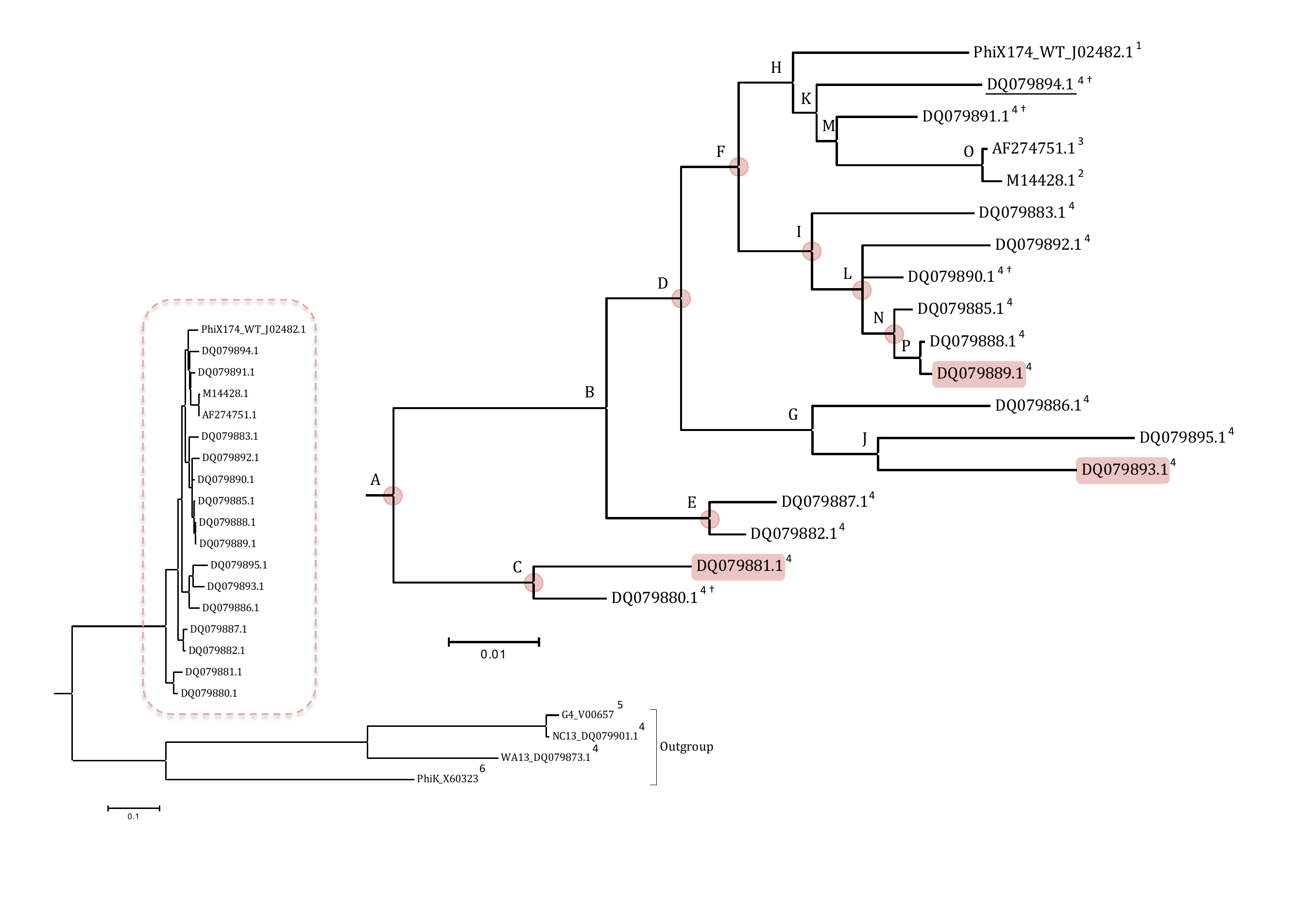}
\caption{Phylogenetic tree of the coat protein of the \phix \ and related phages used in this study. 1: \citet{Sanger1977}, 2: \citet{Lau1985}, 3: \citet{Wichman2000} , 4: \citet{Rokyta2006} , 5: \citet{Godson1978}, 6: \citet{Kodaira1996}.
$\dag$ Extant species with a coat protein identical to an ancestral haplotype. Underlined Species has a coat protein identical to the consensus of extant sequences.
Nodes marked from A to P represent the internal nodes for which the ancestral reconstructions were performed, see also Table \ref{Table:Polymorphisms}. Highlighted extant species and nodes with bullets indicate presence of epistatic interactions.}
\label{Fig:PhylogeneticTree}
\end{figure}

\section{Results}

\subsection{Phylogeny and ancestral reconstruction}
\begin{figure}[t]
\includegraphics[width=\textwidth]{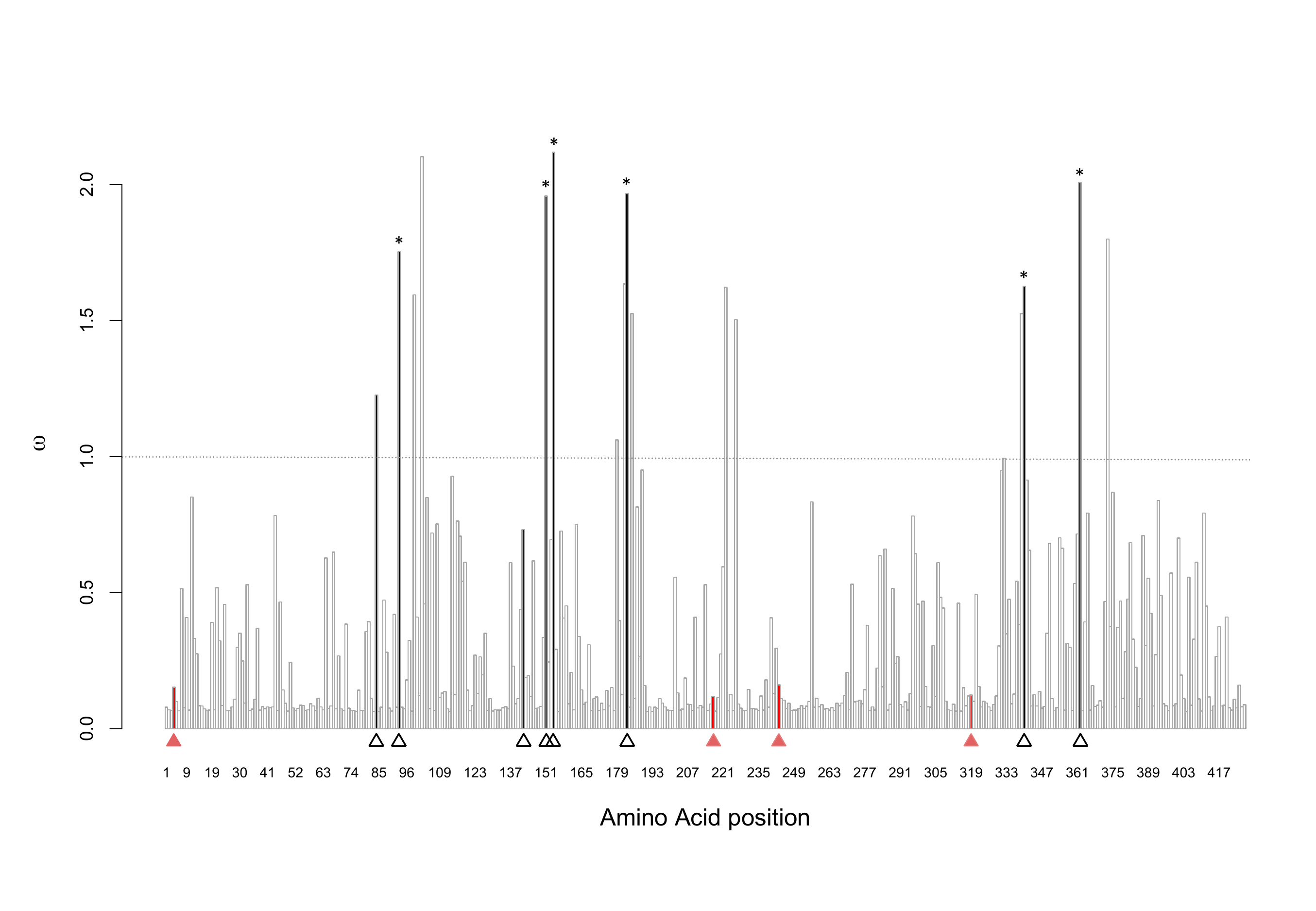}
\caption{Distribution of $\omega$ $(dN/dS)$ per amino acid site of the \phix \, coat protein estimated by PAML, using the model M8 and the consensus tree from the Bayesian analyses (Fig. {\ref{Fig:PhylogeneticTree}},and Materials and Methods section). The mean value is $\overline{\omega} = 0.084$. The ancestrally fixed sites are indicated by filled triangles (red online), and the ancestral polymorphisms by open (black) triangles. * Ancestral polymorphic sites with statistically significant $\omega$ value.}
\label{Fig:OmegaDist}
\end{figure}

The Bayesian phylogenetic reconstruction, using a codon model of substitution, can be seen in Figure \ref{Fig:PhylogeneticTree} trees based on nucleotide and amino-acid models showed similar topology, see Methods and Appendix \ref{SI:Phylogenetics}. The average nucleotide divergence among the ingroup \phix \, sequences was less than 5\%, with a low average $dN/dS$ ratio $\omega = 0.060$ (95\% Credible interval: 0.049, 0.092) in MrBayes \citep{Ronquist2012} and a maximum likelihood estimate of $\omega = 0.084$ using PAML \citep{Yang:2007aa}. Both measures are consistent with strong purifying selection (see text and Fig. \ref{Fig:OmegaDist}).

The ancestral reconstruction for the ingroup \phix \, shows 34 nucleotide differences when compared with the WT \phix . Of these, 12 are non-synonymous and 22 synonymous. Because our goal is to study the energetic changes in the major coat protein, we focus our analyses on the non-synonymous changes. Table \ref{Table:Polymorphisms} and Fig. \ref{Fig:PhylogeneticTree} summarize the number of haplotypes and substitutions for each node on the tree.

From the 12 non-synonymous changes in the ancestral node 8 can occur as two different alleles. We refer to the later as ancestral polymorphisms. Considering all possible combinations of these allelic pairs results in 256 putative ancestral haplotypes. The four remaining positions (3V, 216R, 242F, 318R) are fixed in the ancestral and internal nodes as well as in all extant species except the WT, and will be referred to as Ôancestrally fixedÕ.

For simplicity, in this paper, the ancestral haplotype containing this four fixed positions plus the remaining eight in the same state as the canonical WT, is dubbed Ancestral Wild Type (AWT, see Table \ref{Table:Polymorphisms}); the ancestral haplotype containing all 8 polymorphic positions in a state different from the AWT is called AWT$^{(8)}$. AA substitutions will always be reported relative to the AWT; for example, T92S means that the AWT has a threonine (T) at position 92, which is substituted by a serine (S).

The per site distribution of $\omega$ (Fig. \ref{Fig:OmegaDist}) in the coat protein reveals that most of the sites in the protein have low $\omega$ values; this includes the four ancestrally fixed positions. This is a typical pattern resulting from purifying selection \citep{Goldman1994}. Of the eight variable sites in the ancestral node, six are under diversifying selection, indicated by a statistically significant value of $\omega > 1$ (Fig. \ref{Fig:OmegaDist}). 
Interestingly, the consensus sequence of all extant haplotypes coincides with one of the ancestral haplotypes (Q153E). Moreover, this same haplotype also corresponds to the AA sequence of the coat protein of the extant species DQ079894.1.

We also found that three other ancestral haplotypes are also present in the extant species, namely DQ079890.1, DQ079891.1, DQ079880.1 (Fig. \ref{Fig:PhylogeneticTree}).

We stress that the set of ancestral haplotypes should not be interpreted as a biological population; it is only the set of all probable ancestors inferred using Bayesian methods. Each ancestral haplotype has a certain probability of being the true ancestor. In our case, the most likely haplotype has only 3 differences from the AWT (T92S, Q153E and S339A; see Table \ref{Table:Polymorphisms}), and has a probability of $\Pr= 0.067$. This is much larger than the uninformative prior probability, where each ancestor has $\Pr= 1/256 \simeq 0.0039$.

\subsection{Spatial location of the ancestral haplotypes.}

\begin{figure}[t]
\includegraphics[width=\textwidth]{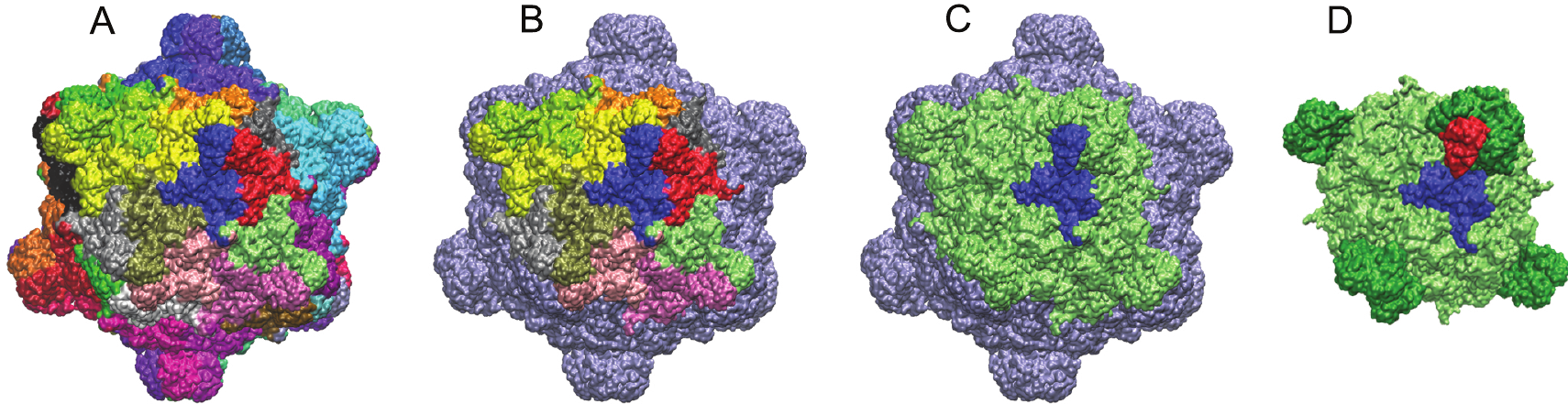}
\caption{Capsid structure and molecular model of the fragment. Molecular model of the \phix \, capsid  highlighting (A) the repetitive constituting protein subunits (each color is a subunit) , (B) the 12 proteins subunits in the fragment employed for our calculations, (C) the fragment and one protein subunit and (D) the fragment showing one coat proteins (blue) and one spike proteins (red). }
\label{Fig:CapsidModel}
\end{figure}

\begin{figure}[p]
\centering
\includegraphics[scale=0.3]{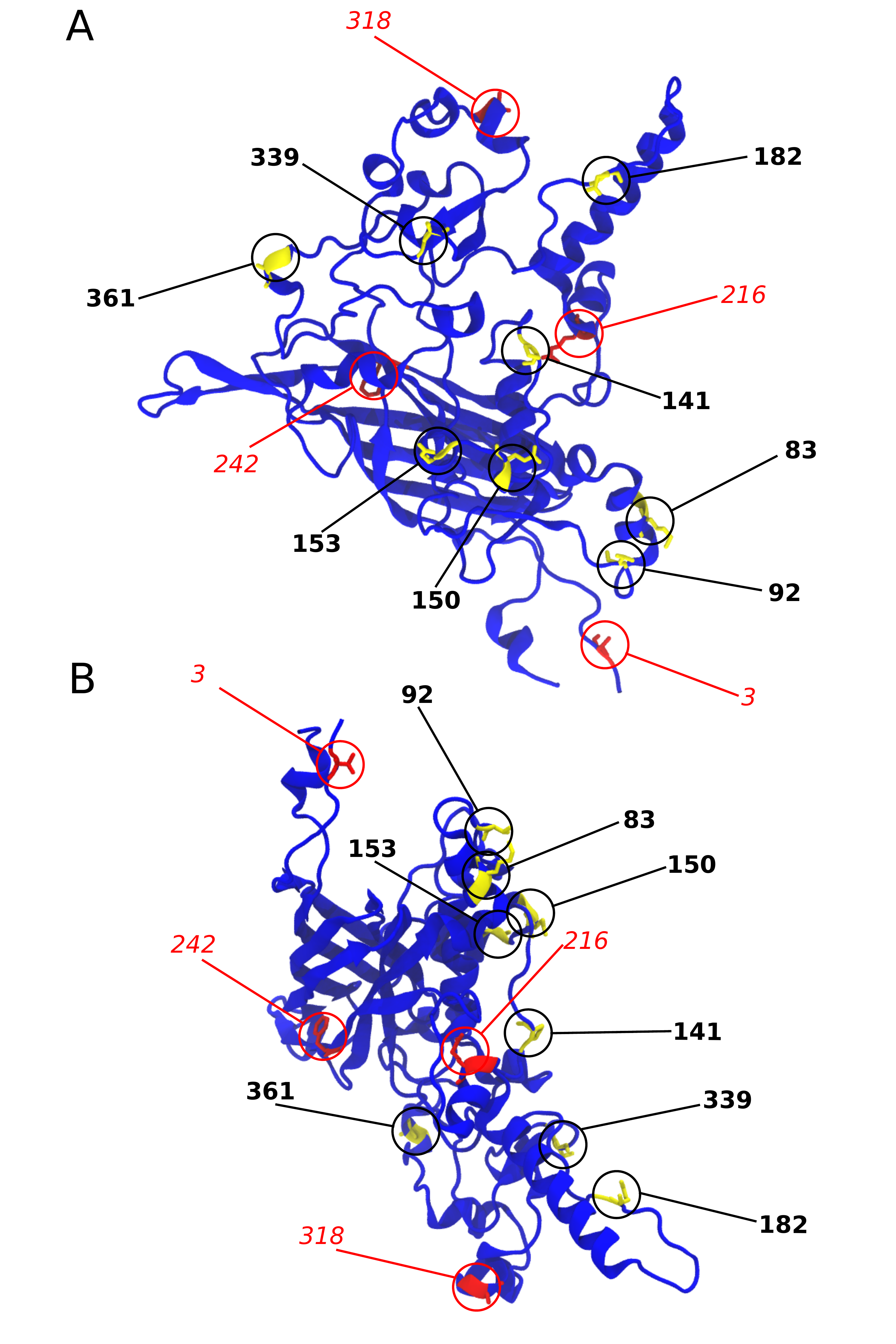}
\includegraphics[scale=0.5]{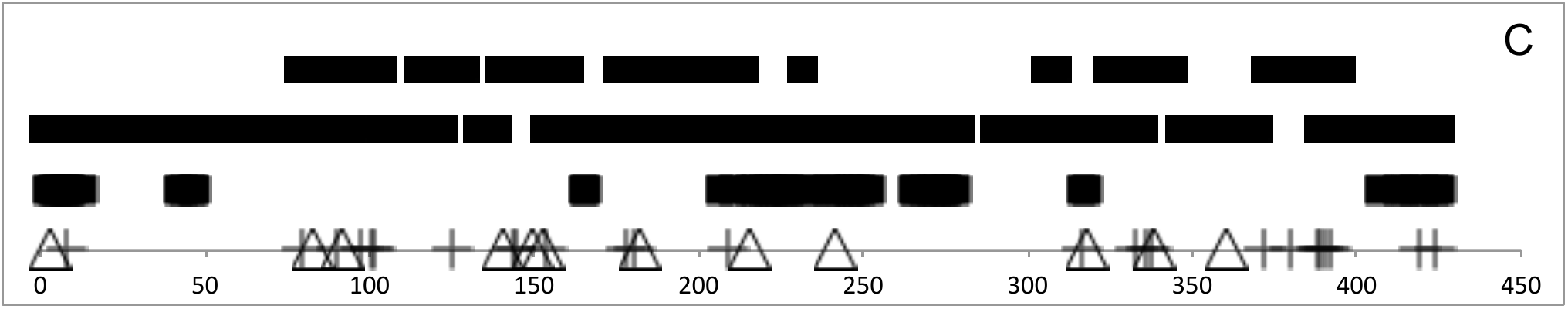}
\caption{Coat protein and spatial location of the variable sites. (A) Top view (from the solvent's perspective). (B) Side view. The residues in yellow pointed by black bold italic labels are  the ancestral polymorphic  sites. The residues in red, pointed by red upright regular fonts are the ancestrally fixed sites. (C) Classification of the position of the amino acids according whether they are exposed to the solvent (top bars), are at the interface between proteins (middle bar) or are exposed to the internal space of the capsid (lower bar). Crosses: variable sites in the extant species; triangles: ancestral variable sites.}
\label{Fig:PaprikasAndBananas}
\end{figure}

Figure \ref{Fig:CapsidModel} shows the crystal structure of the WT \phix, and details the fragment that we employed for our calculations. Figure \ref{Fig:PaprikasAndBananas}A,B shows the position of the ancestral and extant variable sites in the coat protein. Most of the polymorphic sites face the external milieux, suggesting that these are relatively free of structural constraints (Fig.  \ref{Fig:PaprikasAndBananas}C), and the ancestrally fixed sites are mostly facing the inside. Most of the ancestral polymorphic AAs are close enough as to be able to interact electrostatically, although that is contingent on the specific substitutions.

\subsection{Distribution of free energy at the ancestral state and in extant species}
Through in silico structural modeling we calculated the free energy of each possible haplotype at all internal nodes and extant species. For this purpose we use the software FoldX, which gives an estimate of the free energy changes, \ddg , of the haplotypes relative to a known structure (see Materials and Methods). We chose the AWT as a reference, because it has the least possible changes (4 AAs) from the known structure of the capsid of WT \phix . In this way we compare the free energy of the derived species (internal nodes and extant species) to an evolutionary equidistant reference point at the ancestral state and not to an extant leaf (e.g. the WT), which can also avoid potential biases in our analyses.

\begin{figure}[t]
\centering
\includegraphics[scale=0.35]{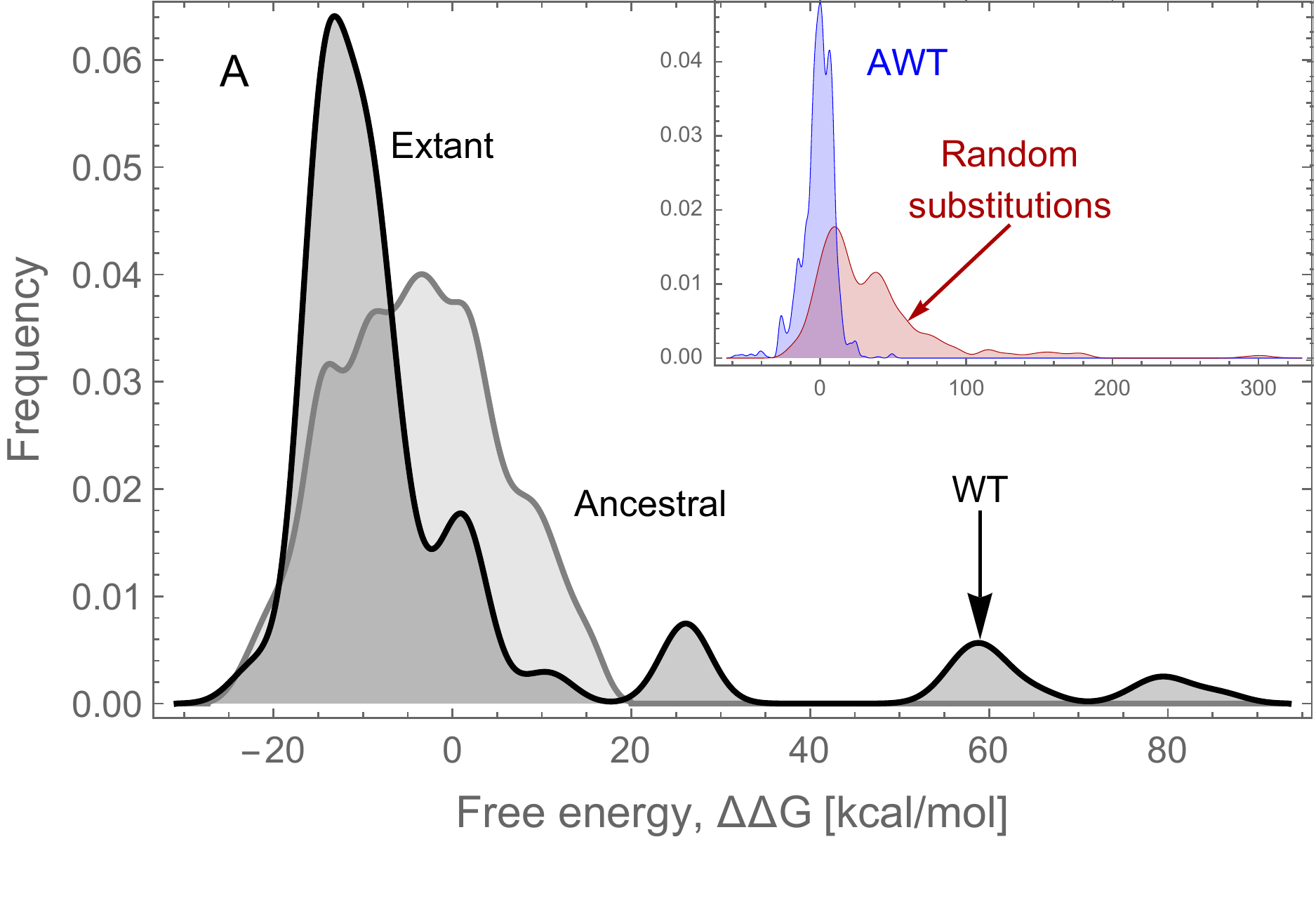}
\includegraphics[scale=0.35]{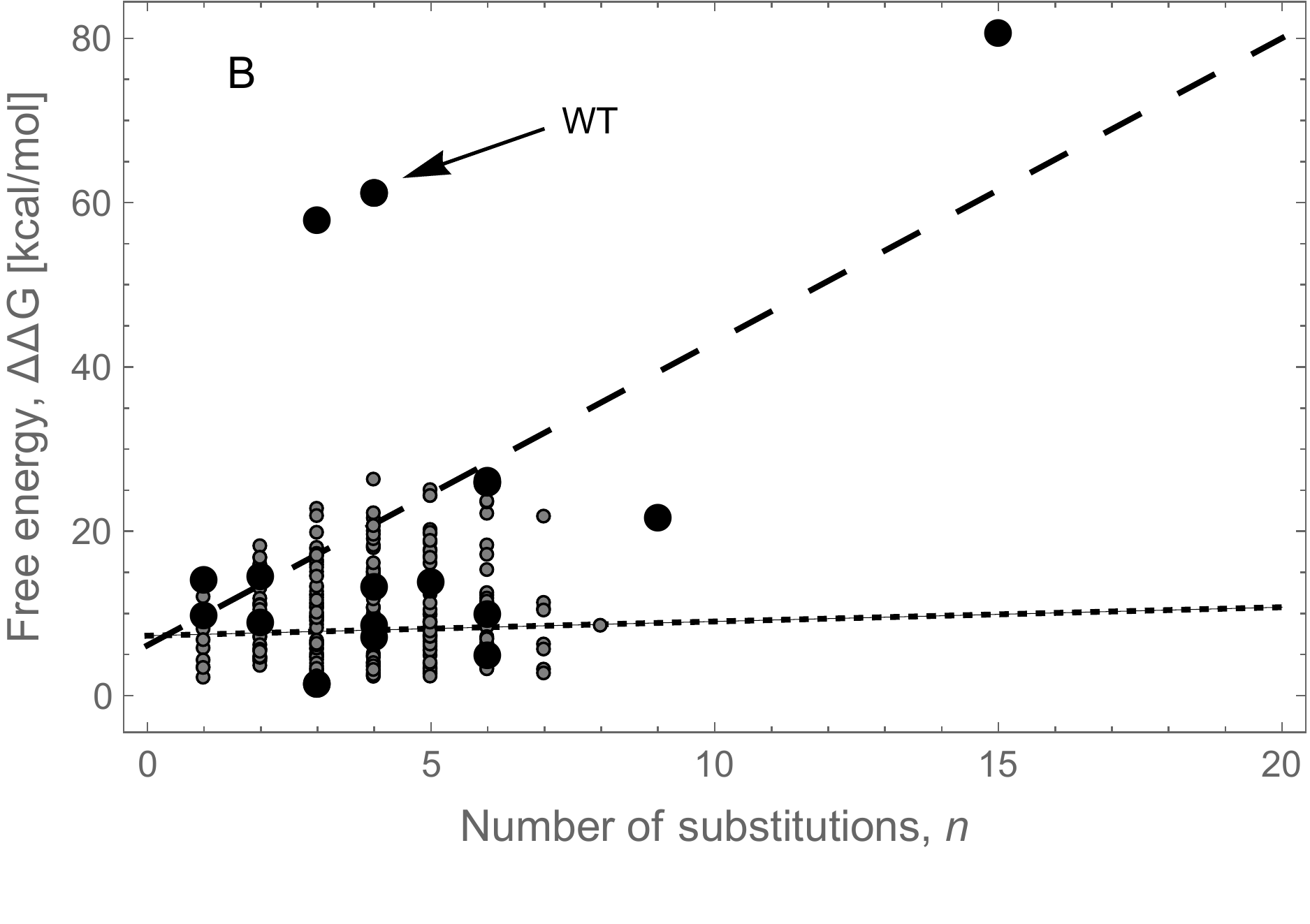}
\caption{(A) Histogram of free energies relative to the AWT. Black: extant species (mean=7.89 kcal/mol), gray: ancestral haplotypes (mean=-3.53 kcal/mol). Inset: histogram of a set of 127 random mutants having between 1 and 5 substitutions. (B) Relationship between free energy (relative to the AWT) and the number of amino acid substitutions ($n$). Black large bullets: extant species; gray small bullets: ancestral haplotypes. The lines are linear regressions. For the ancestrals (dotted gray line): \(\Delta \Delta G =  7.28+0.17 n\) [kcal/mol] ($p=0.49$); for the extant species (dashed black line): \(\Delta \Delta G =  6.11+3.70 n\) [kcal/mol] ($p=0.024$).}
\label{Fig:FreeEnergyDist}
\end{figure}

The free energy amongst all putative ancestors has an average difference of $\Delta \Delta G=-3.53$ kcal/mol, which is significantly different from zero ($p\sim 10^{-9}$). We find no significant difference between the mean of the free energies at the ancestral node and at the tips (Mann-Whittney test, $p=0.30$; Fig. \ref{Fig:FreeEnergyDist}A). Free energy at the ancestral state is normally distributed with mean as indicated above with a variance of 80. The extant species have a mean $\Delta \Delta G=7.89$ kcal/mol, with a variance of 915; the latter is significantly larger than that of the ancestrals (Conover variance ratio test, $p=0.021$). 

Because the ancestral haplotypes are all putative ancestors, the statistics above only reflect the uncertainty of the free energy distribution at the ancestral state, not an estimate of the standing variation of the ancestral population.

\subsection{The energy spectrum of random substitutions is wide}
Randomly mutating the capsid gene results on a spectrum of free energy that is much wider than that of the ancestral or extant species (Fig. \ref{Fig:FreeEnergyDist}A). While the maximum \ddg \, in the ancestral haplotypes is about 24 kcal/mol, in the set of random substitutions it is 300 kcal/mol. Curiously, both of these extreme values occur with only four substitutions (Fig. \ref{Fig:FreeEnergyDist}A). The smaller variance of the ancestral distribution is consistent with the hypothesis that purifying selection maintains the capsid, provided that mutations resulting in large free energy deviations are unfit.

\subsection{Mutational effects have a positively skewed distribution}
Following the terminology from quantitative genetics \cite[][p. 122; see also \citet{Crow:2010gh,Cordell:2002vo}]{Falconer:1981} we refer to the free energy difference of a single substitution as the additive mutational effect. These effects have a mean value of 5.43 kcal/mol per AA and are described by a skew normal distribution with the positive shape parameter, indicating a bias towards positive values (Fig. \ref{Fig:AminoAcidEffects}). While some substitutions can have effects as small as 0.079 kcal/mol, we also find that some can be as high as 41.84 kcal/mol (in absolute value). This is consistent with the shape of the distribution of allelic effects of quantitative traits, which show positive skews \citep{Keightley:2007hq,EyreWalker:2007dl}.

\subsection{Average increase of free energy with substitution number}
Although we expect the free energy to increase with the number of substitutions, a linear regression (Fig. \ref{Fig:FreeEnergyDist}B) indicates that in the ancestral set there is no significant trend (slope=0.29, $p=0.25$). This robustness of the free energy distribution against number of substitutions is expected when we take into account that the AWT is an arbitrary reference sequence and as such it should not reveal any informative pattern regarding differences from it. 

We find the opposite behavior in the extant haplotypes: there is a marked trend, where each substitution adds on average 3.70 kcal/mol ($p=0.020$). This slope is consistent with the mean value of the distribution of mutational effects. The difference in the trends between ancestral and extant can be explained by mutation accumulation along the evolution of the capsid. We note that this trend is heavily driven by three points which have notably high free energies (DQ079885, WT and DQ079892; the latter having 15 substitutions). However, even if we remove these points we still find a positive trend (slope=1.60, $p=0.085$).

\subsection{Few substitutions drive free energy changes}
Relative to the AWT, the WT coat protein has a larger free energy of $\Delta \Delta G=61.24$ kcal/mol ($p \sim 10^{-26}$). Two substitutions, R216H and F242L, contribute by 64.07 kcal/mol to this energetic deviation and the two remaining substitutions (V3I and A318V) together contribute with $-4.41$ kcal/mol. Although none of these four substitutions change AA properties (charge or hydrophobicity), the former two involve aromatic rings which can account for their large energetic effect due to the   steric reconfigurations of the local molecular environment and/or by changing the nature of the van der Waals interactions.

Species DQ079885 has a similarly large deviation relative to the AWT ($\Delta \Delta G=58$ kcal/mol), again most of it is due to two substitutions (D338H and E145D) adding up to 62 kcal/mol; the third substitution, S339A reduces it by 4.60 kcal/mol. (Note the common presence of histidine). Of the 15 mutations of species DQ079892, three (Y102S, Q154N and D338N) contribute with 70.80 kcal/mol (88\%) of the $\Delta \Delta G=81$ kcal/mol. The remaining 12 are not all small, but additively compensate for the remaining 12\%. A central conclusion from these examples is that that most of the deviations in free energy are driven by only a few substitutions.

\begin{figure}[t]
\centering
\includegraphics[scale=0.35]{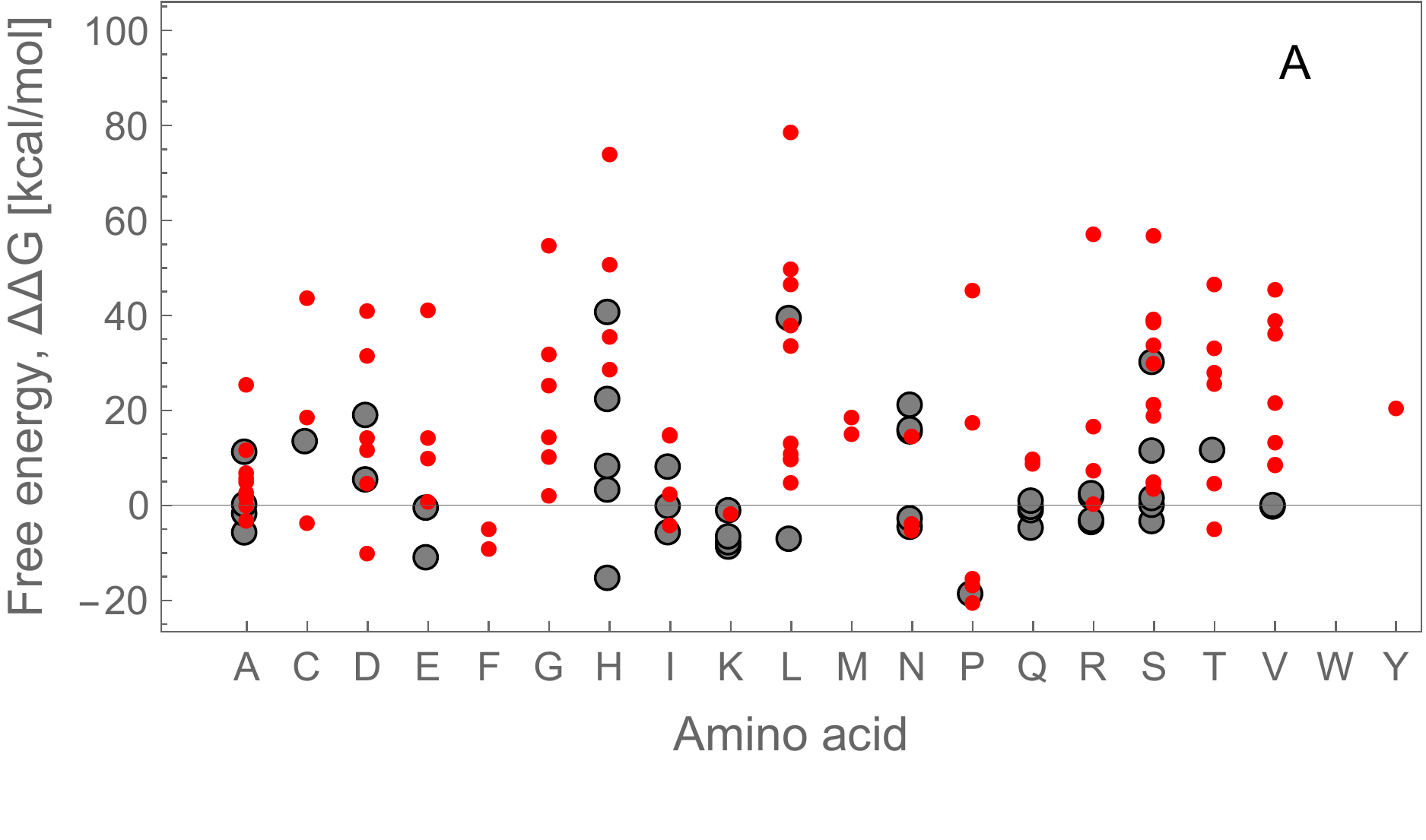}
\includegraphics[scale=0.35]{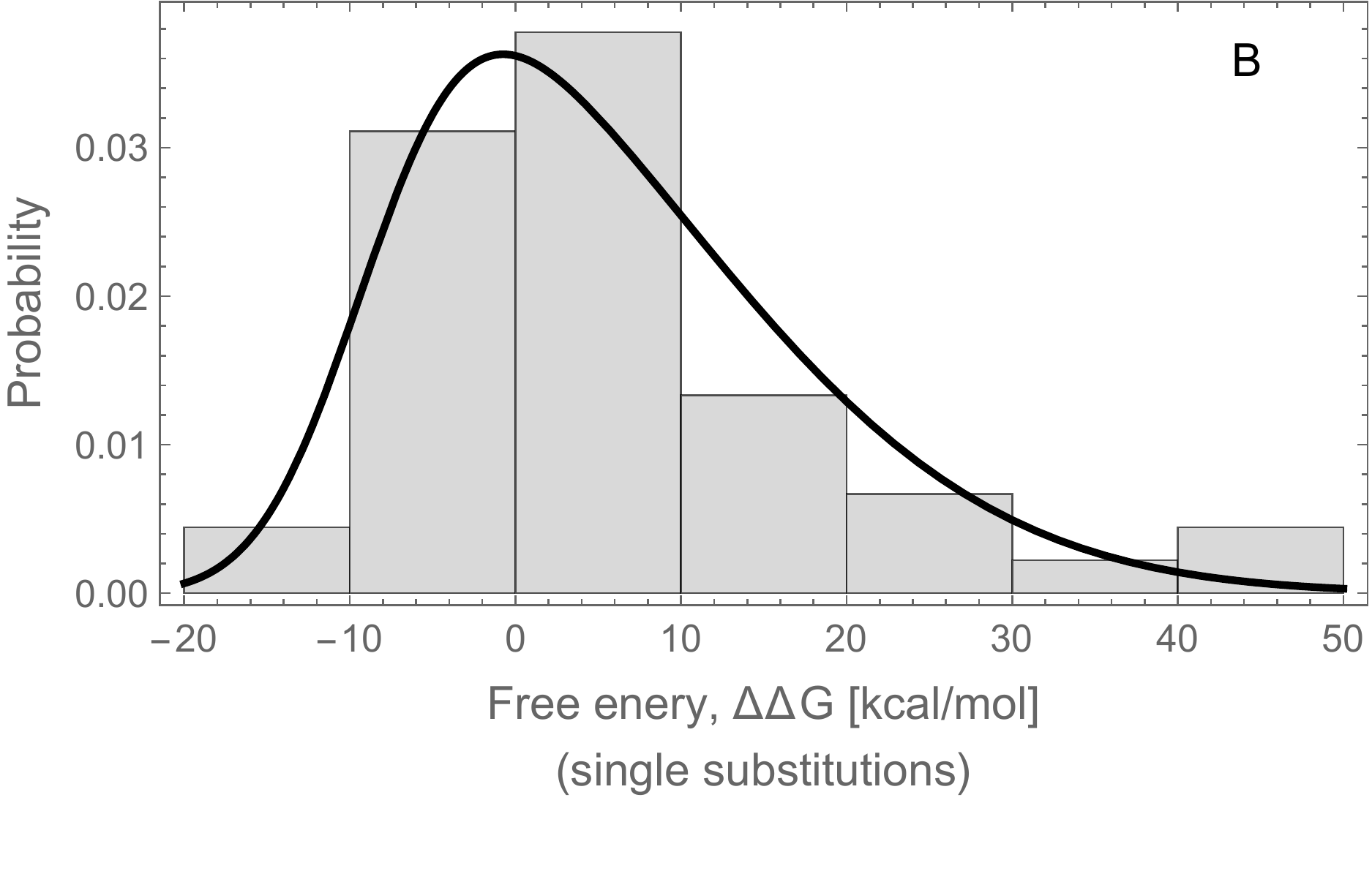}
\caption{(A) Energy contributions of the different amino acids (based on single substitutions to the AWT). Large (gray) bullets: single substitutions occurring in the alignment; small (red, color online) bullets: single substitutions of a randomly generated data set. (B) Distribution of single-substitution effects (only of the single polymorphisms in the alignment). The solid curve is a skew-normal distribution with location, scale and shape parameters $-9.02,18.80, 3.57$ (maximum likelihood estimators), respectively (mean=5.43, variance=144.82 and skewness=0.75).}
\label{Fig:AminoAcidEffects}
\end{figure}

	The free energy differences of every single substitution varies according to three factors: the original and derived AAs, and the position in which these occur (Fig. \ref{Fig:AminoAcidEffects}). Notably, histidine and leucine tend to have the strongest effects. The distribution of effects is heavily skewed, biased towards increasing the free energy, even in the set of random substitutions (Fig. \ref{Fig:FreeEnergyDist}). This asymmetry is suggestive of the capsid being close to an energetic minimum.

\subsection{Strength and causes of epistasis}
\begin{figure}[t]
\centering
\includegraphics[scale=0.35]{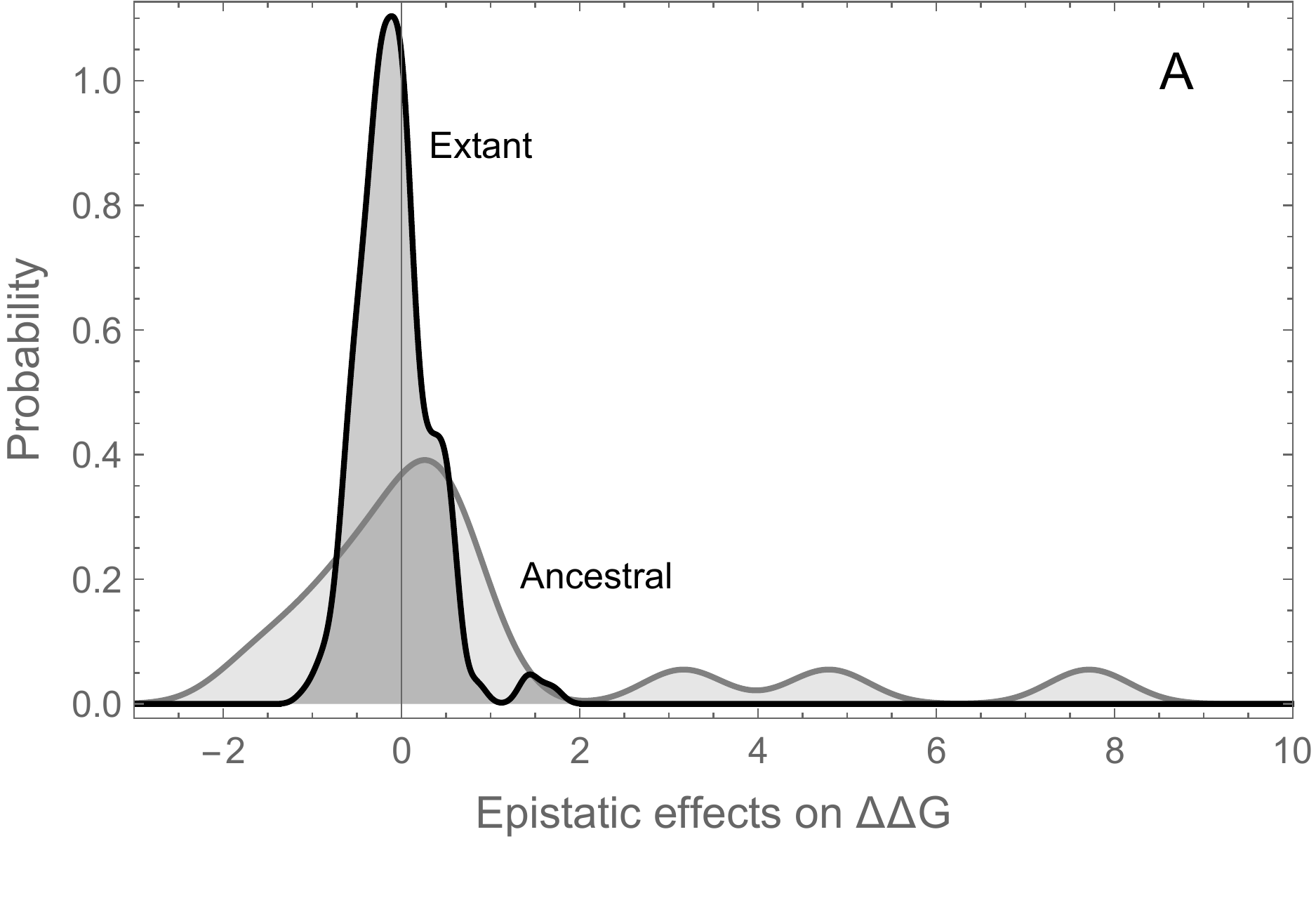}
\includegraphics[scale=0.35]{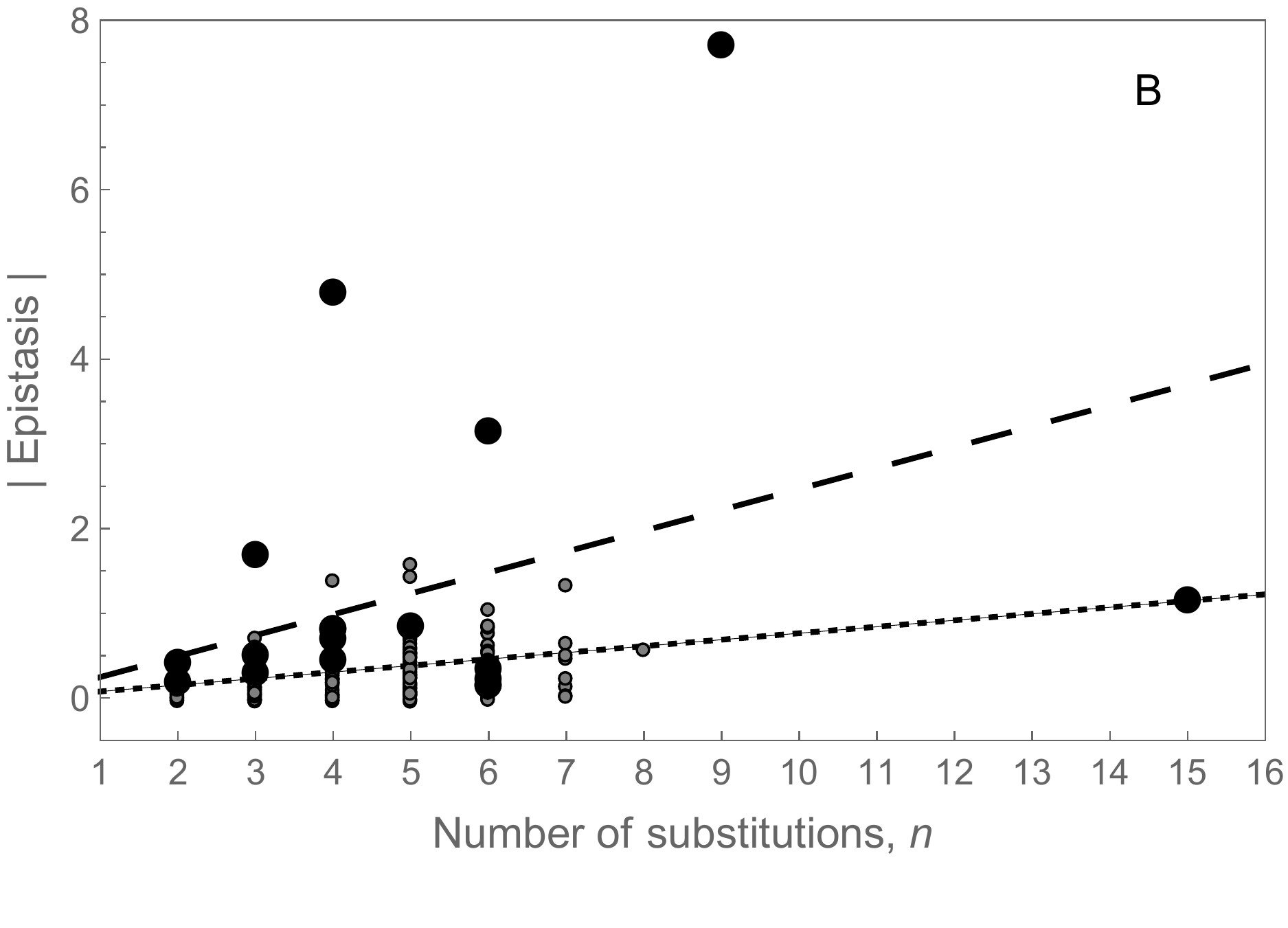}
\caption{
(A) Histograms of epistatic values. Black: extant species (mean=-0.87 kcal/mol; variance=5.84), gray: ancestral haplotypes (mean=-0.071 kcal/mol; variance=0.18).  (B) Relationship between epistasis and the number of amino acid substitutions ($n$). Black large bullets: extant species; gray small bullets: ancestral haplotypes. The lines are linear regressions. For the ancestrals (dotted grey line): \(\epsilon =  0.075  n\) [kcal/mol] ($p=0.00$); for the extant species (dashed black line): \(\epsilon=  0.25 n\) [kcal/mol] ($p=0.0029$).}
\label{Fig:EpistasisDistribution}
\end{figure}

Structural epistasis is estimated by comparing the free energy of each haplotype with the additive free energy of the constituting single substitutions. We declare a haplotype to be epistatic ($\epsilon\neq 0$) when the $p$-value of a tailored T-test is equal or less than a threshold value $p^*$. The threshold depends on the specific haplotype when considering the Bonferroni correction, and is roughly, $p^* \simeq 2\times 10^{-3}$ (see Materials and Methods and Appendix \ref{SI:Bonferroni}). Although for both the ancestral haplotypes and the extant species the average value of the distribution of epistasis is low (Fig. \ref{Fig:EpistasisDistribution}), they are significantly different from each other ($p=0.00$).

We identified 38 epistatic haplotypes ($\sim$15\% of the sequences with more than one substitution). Of these, 34 are in the ancestral set, 3 in internal nodes (1 in Node C \textendash \ K83Q, T92S, G101R, P141A, Q153E, Q182L, S339A and 2 haplotypes \textendash \ E150Q, S339A and E150Q, Q153E, S339A \textendash \ also shared among several ancestral nodes), and 3 in the extant species (DQ079889, DQ079893, DQ079881; Fig. \ref{Fig:PhylogeneticTree}).

In the ancestral set a linear regression of epistasis on the number of substitutions results in a poor trend (slope=0.075 kcal/mol). However, the trend is strong on the extant species, with a significant slope of 0.25 kcal/mol  (Fig. \ref{Fig:EpistasisDistribution}B) .

The three extant species that show strong epistasis are DQ079889, DQ079893 and DQ079881. None of these have large energetic deviations. For species DQ079889, DQ079893 epistasis is positive, namely of the same sign as the additive effects while species DQ079881 has negative epistasis, providing in this case a moderate compensatory effect.

\subsection{Molecular interactions account for structural epistasis}
\begin{figure}[t]
\includegraphics[width=\columnwidth]{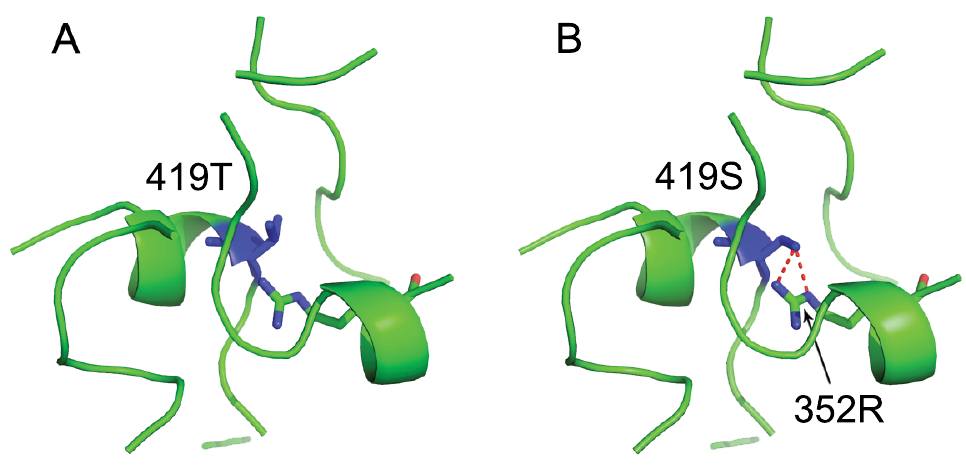}
\caption{The amino acid substitution T419S results in hydrogen bonding. (A) Detail of a non-epistatic haplotype in node P (see Fig. \ref{Fig:PhylogeneticTree}) where (B) the derived amino acid 419S in species DQ079889 can bond with 352R (red dotted lines; color online). This local changes result on epistasis not through the hydrogen bond, but by disrupting the local energy field which propagates to other substitutions.}
\label{Fig:AAInteraction}
\end{figure}

Our tenet is that epistasis is correlated with number of substitutions due to structural reasons. Larger numbers of substitutions make it more likely the derived AAs result spatially close, and thus able to change the nature of their electrostatic interactions. Yet, in the extant species presence/absence of epistasis can be associated with single substitutions. We illustrate our point by analysing in detail the AA interactions of the three haplotypes named above, which have the largest epistasis. For species DQ079889, the side chains of one of the four substitutions, T419S makes the serine side chain able to interact with  352R (the side chains are spatially close, separated by about 2.64 \AA , and can make a hydrogen bond;  Fig. \ref{Fig:AAInteraction}).  A similar case can be made with the substitution E150A from node J to species DQ079893 (showing epistasis): 150A can newly establish hydrogen bond with 146A and 147N whereas 150E cannot. We also find that in the substitution Q380P from node C to species DQ079881, the derived AA 380P establishes a hydrogen bond, this time not with another AA, but with the backbone. This interaction still results in an increase in epistasis.

Although in each case the presence of epistasis can be associated with a specific  AA, all these species have at least one more substitution (relative to their parental node). This means that epistasis is not generally  resulting from hydrogen bonds amongst substitutions. Instead, hydrogen bonds modify the local molecular environment resulting in changes of the energy field that propagate beyond the immediate radius of the interactions. Hence, epistasis is not only mediated by hydrogen bonds, but can also be determined by other electrostatic factors such as Coulomb and van der Waals interactions (Table \ref{Table:HydrogenBonds}).

\begin{table}[t]
\centering
%\rowcolors{2}{light-gray}{white}
\begin{tabular}{ccclc}
\hline \hline
Node								&	species						& H-bond	&	Subst.		&	Epist.		\\ \hline
\rowcolor{lightgray} 					&	DQ079880					&	-1		&	R101G			& 	L		\\
\rowcolor{lightgray} \multirow{-2}{*}{\it\underline{C}}		
									&	{\it\underline{DQ079881}}	&	-1		&	Q380P			& 	M	\\ 
									&	DQ079882					&	0		&	T92A				& 	L		\\
\multirow{-2}{*}{\it\underline{E}}		&	DQ079887					&	+1		&	T92N			& 	L		\\
\rowcolor{lightgray}					&	{\it\underline{DQ079893}}	&	-2		&	E150A			& 	G		\\
\rowcolor{lightgray} \multirow{-2}{*}{J}		&	DQ079895			&	+1		&	Q125R			& 	-			\\ 
									&	DQ079888					&	0		&	E150A			& 	-			\\
\multirow{-2}{*}{P}					&	{\it\underline{DQ079889}}	&	+1		&	T419S			& 	G		\\ 
\hline \hline 	
\end{tabular}
\caption{Change in the number of hydrogen bonds in the substitutions leading to some extant species (as in Fig. \ref{Fig:PhylogeneticTree}). Hydrogen bonds are define based on geometric criteria: if two atoms which can form hydrogen bonds are close enough we assume that these are created among the amino acids in question. Underlined italic symbols: epistatic haplotypes. G: gains epistasis; L: losses epistasis; M: maintains epistasis}
\label{Table:HydrogenBonds}
\end{table}

\subsection{Statistical and structural epistasis are correlated but not causally}
Structural epistatic effects, as defined here, are intimately related to interaction energies amongst AA side chains. Thus to make a stronger case for evolutionary biology we also estimate statistical epistasis employing an ANOVA on the free energy data. For this purpose, we use another independent set of replicas of the 256 ancestors.

\begin{figure}[t]
\centering
\includegraphics[width=\columnwidth]{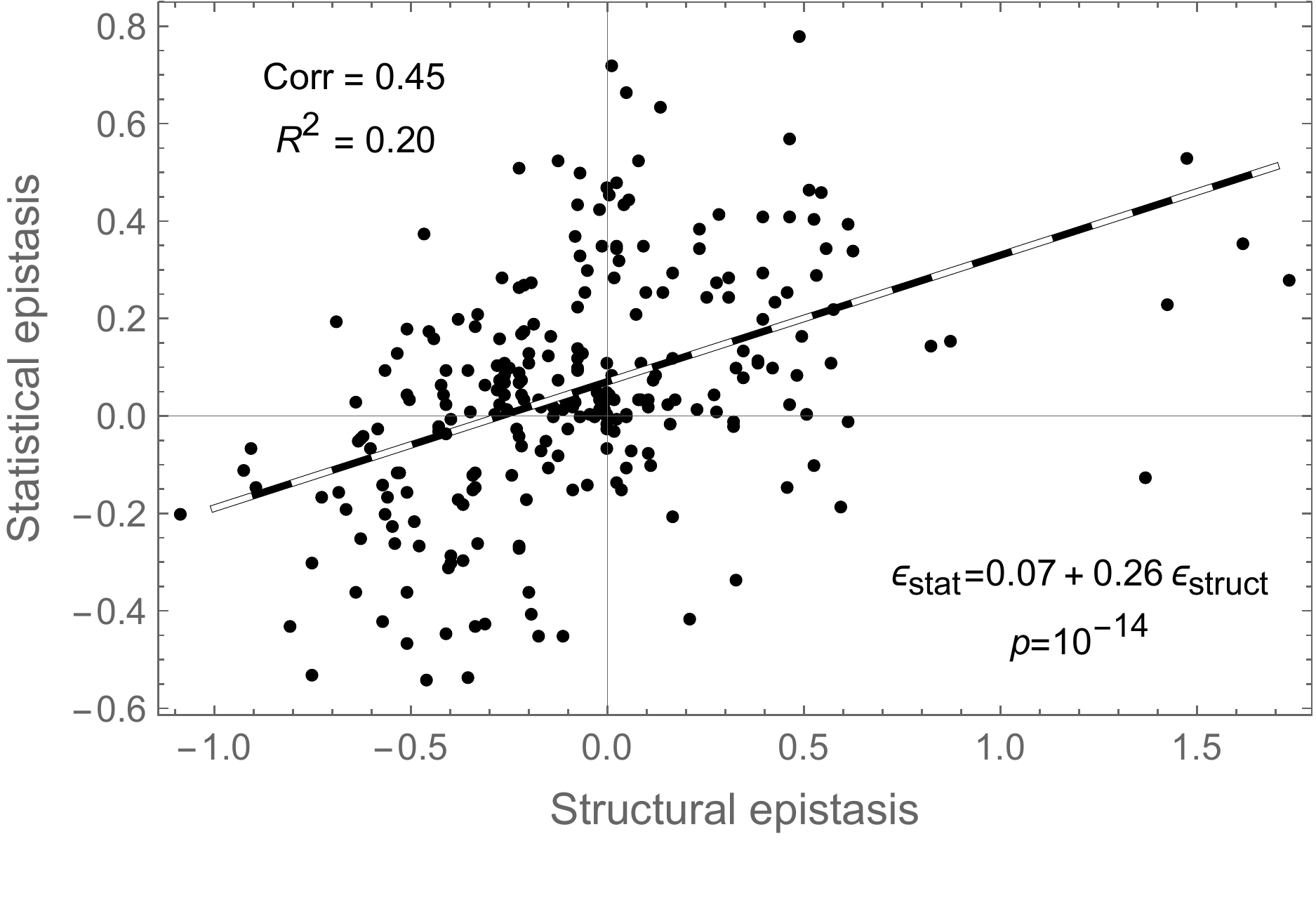}
\caption{Correlation between statistical and structural epistasis (see text for the definitions). The data used for estimating each type of epistasis are from two sets of independent simulations of the 256 ancestral haplotypes. The dashed line is a linear regression between the two measures. Each point is an average over at least 15 replicas.}
\label{Fig:StatVsStructEpistasis}
\end{figure}

The usage of ANOVA and related methods is typical in quantitative genetics to estimate epistasis. We tested six models considering from additive and up to 5-way mixed effects (see Methods). The model with up to 5-way effects is preferred on the basis of Akaike's information criterion, even when it has 219 parameters. However, only 19 of these are significant (with confidence $p=0.050$). Nevertheless, statistical epistasis correlates with structural epistasis ($p \sim 10^{-14}$; Fig. \ref{Fig:StatVsStructEpistasis}); this holds true even if we employ only pairwise regression coefficients (data not shown). In both the pairwise and 5-way effects models, there are 6 of 27 pairwise parameters that are significant. The significant pairs involve two focal amino acids: 83 and 153 (inter-AA distance=33.41\AA). Besides correlating with each other, 83 correlates with 141 (13.47\AA), 150(21.8\AA) and 361(40.1\AA), and 153 correlates with 150(42.86\AA) and 361(13.04\AA). However, none of these pairs show significant structural epistasis. Still, 35 of the 38 epistatic haplotypes involve at least one substitution at positions 83, 150, 153 or 361. Note that although the inter-AA distances will not always allow hydrogen bonds, they are short enough so that substitutions can affect the local energy field.

%When we see the spatial arrangement of these loci, we find that they are not close enough as to interact (Fig. \ref{Fig:PaprikasAndBananas}). However, these positions are faced closely to another identical protein subunit, resulting proximate in space and  able to interact electrostatically inter-molecularly \hl{(verify this result with Tomek's new data)}. Hence, there is direct evidence that structural factors can lead to epistasis, even though the information captured by statistical methods does not directly reflect physical mechanisms.

We point out that by analysing  which alleles associate with presence/absence of epistasis we fail to predict which structural elements are responsible for epistasis (data not shown). Thus, the analysis of causal effects should be performed at the structural level, and not solely by association.

\subsection{High order epistasis cannot always be decomposed into pairwise epistasis}
The correlation between pairwise and 5th order statistical epistasis is almost perfect (slope=0.97, $p=0.00$, $R^2=0.99$, corr=0.99; data not shown). This suggests that pairwise factors lead epistasis in multiple mutants. However, there are only 6 high order (i.e., $>2$) significant coefficients, of which only two (5-way interactions) involve one and two (out of 10) of the lower pairs of factors. Conversely, the high order terms that result from the composition of the significant pairwise factors are all non-significant. 

\begin{figure}[t]
\centering
\includegraphics[scale=0.35]{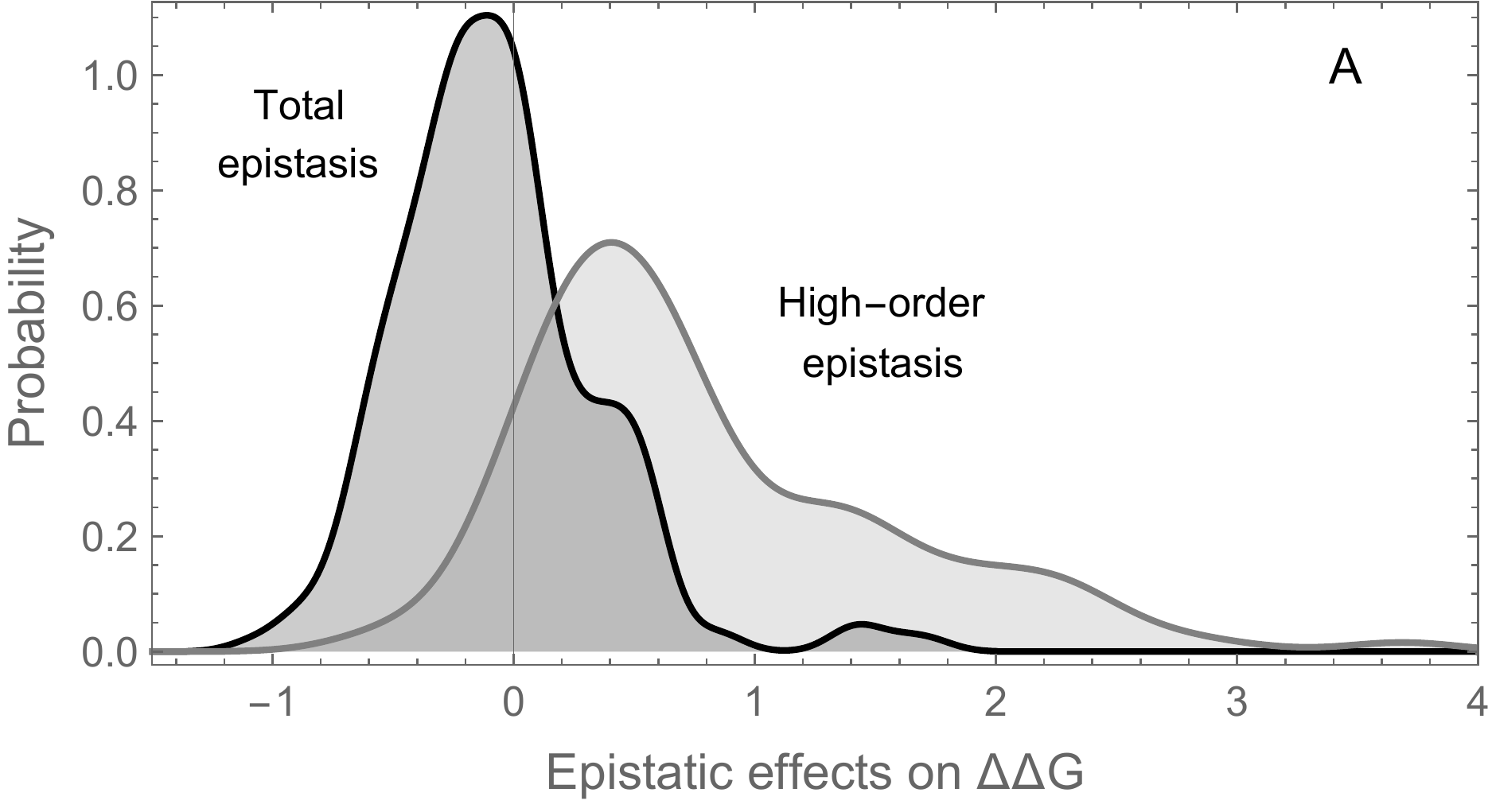}
\includegraphics[scale=0.35]{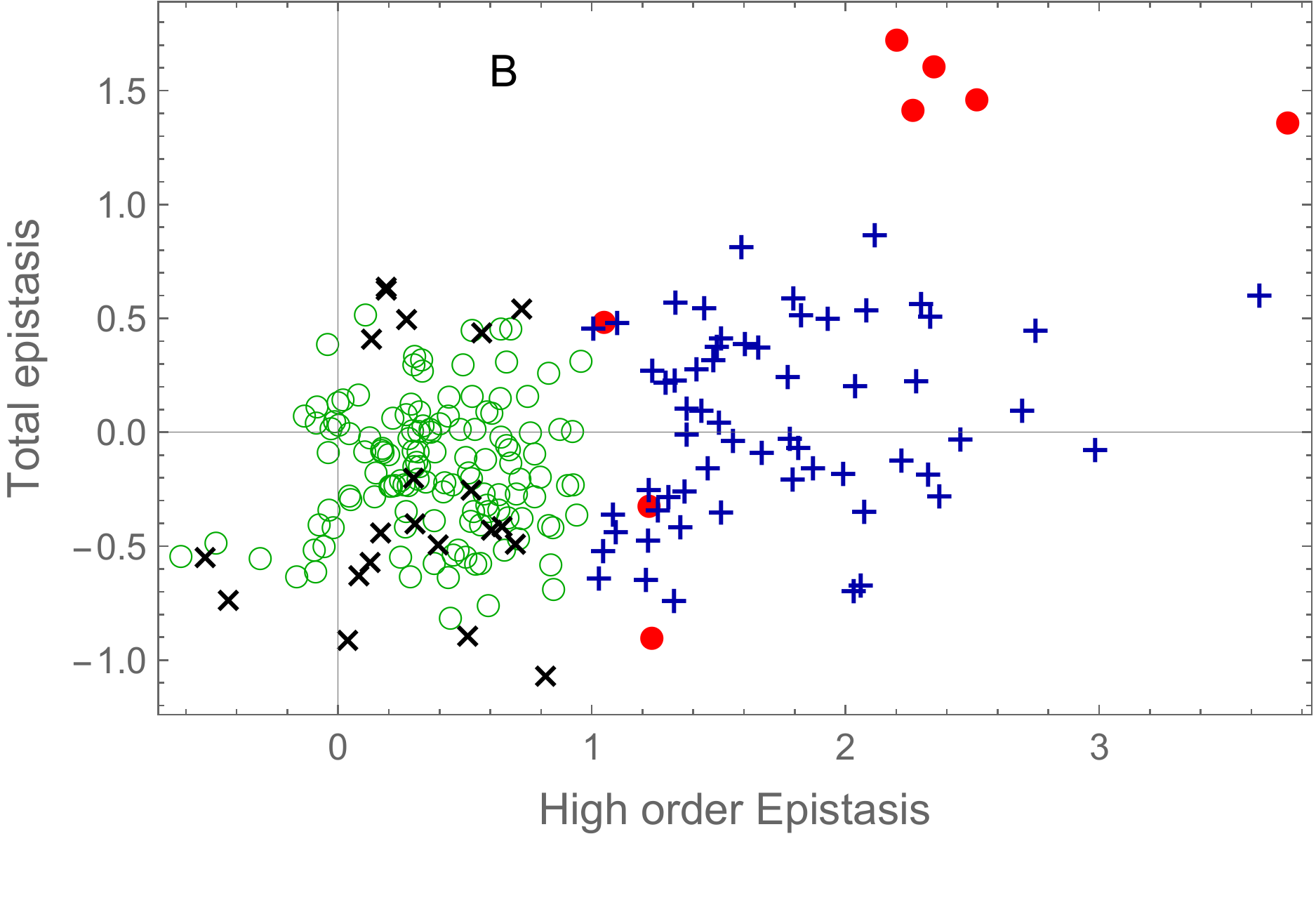}
\caption{
(A) Histograms of total and high order epistatic values. Black: total epistasis in the ancestors  (\emph{idem} as in Fig. \ref{Fig:EpistasisDistribution}A, presented again as a reference), gray: high order epistasis in the ancestors  (mean=0.80 kcal/mol; variance=0.59). (B) Scatter plot between high order and total structural epistasis. (Color online.) Green rings ($\circ$): no epistasis; red bullets ($\bullet$): total and high order epistasis; black times-crosses ($\times$): total epistasis but no high order epistasis; blue plus-crosses ($+$): no total epistasis but with high order epistasis.}
\label{Fig:HighOrderEpistasis}
\end{figure}

We estimate higher order structural epistasis with a similar method with which  estimated total epistasis. Despite the pervasiveness of pairwise effects, there is statistical support for high order epistasis. The distribution of epistasis of high order has a mean of 0.80 kcal/mol, which is significantly different than zero (sign test for the median, $p\sim10^{-9}$; Fig. \ref{Fig:HighOrderEpistasis}A), and its variance is 0.59, which is significantly larger than that of total epistasis (=0.18).
 
There are only 5 epistatic pairs in the set of ancestors: (92, 339), (92, 361), (141, 361), (150, 339), (182, 339). Of the remaining 33 epistatic  haplotypes with three or more substitutions, 28  involve at least one of these epistatic pairs, and in one case all up to four. Although this seems as supporting evidence for the reducibility of multiple way epistasis into pairwise effects, combinatorics of the loci rules it out.

First note that these 5 pairs include in total 6 of the 8 ancestral polymorphisms. Hence, it is highly likely that in subsets of 3 or larger, two or more of the 6 elements appear. (However, this calculation overestimates this probability, because this choice might not include the actual pairs - e.g., (92, 141, 339) includes none of the pairs in the set.) 

In addition to this, we must also consider a structural factor: because there are only a few mutations, on the basis of random incidence, the substitutions are spatially distant. Thus, the occurrence of triplets or higher order interaction structures is unlikely at low mutation rates.

	Moreover, we find that there can be compensatory effects between epistatic terms. That is, the magnitude of epistasis results from the superposition of pairwise effects that are of contrary sign to the value of higher-order interactions. To an extreme, these two terms can balance each other, resulting in values that seem to be additive, but have significant epistasis (blue upright crosses in Fig. \ref{Fig:HighOrderEpistasis}).

\subsection{Fitness assays}
Of the ten synthetic constructs, containing the ancestral coat proteins gene (the eight single ancestral polymorphisms, AWT and AWT$^{(8)}$), all but one (E150Q) were recovered. The recovered haplotypes all presented the same plaque morphology as the WT. Absolute fitness was measured using the growth rate per hour as a proxy (see Material and Methods), and relative fitness was obtained by comparing the absolute fitness of a haplotype against both the AWT and WT. However it is important to mention that the fitness estimated by growth rate is actually a composite trait summing several effects as lag time, burst size and adsorption rate, all contributing to the observed growth of the population. We ignored these factors and limited our measurements to countable plaques.

\begin{figure}[p]
\centering
\includegraphics[scale=0.6]{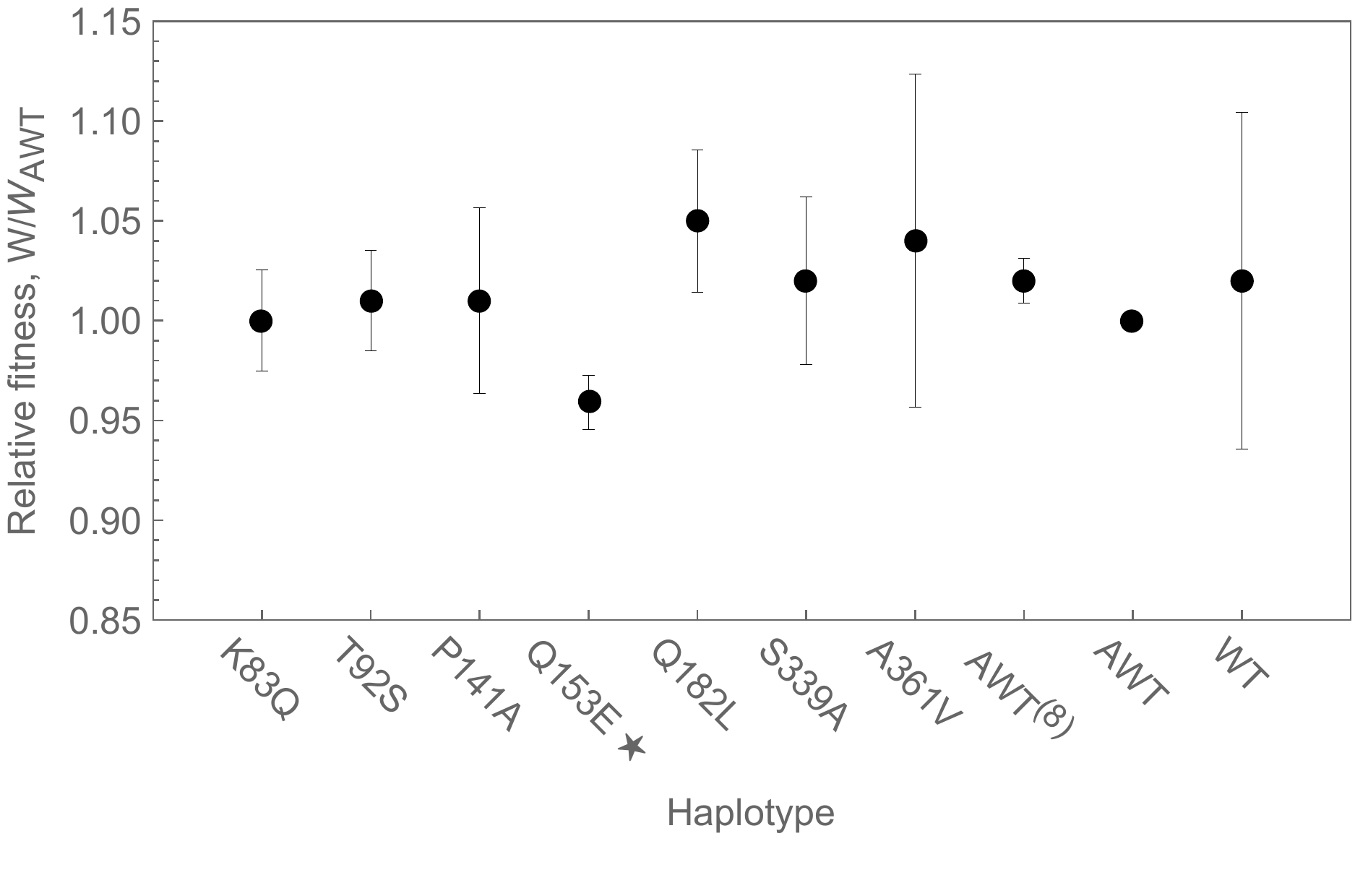}
\caption{Experimental fitness measures (relative to that of the AWT). Bars are on each side one are on standard deviation. None of the haplotypes have significantly different relative fitness from the WT. $\bigstar$ consensus.}
\label{Fig:ExperimentalFitness}
\end{figure}

In order to be consistent with our free energy measurements, which are relative to the AWT, it is convenient to report the fitness relative to the same background (Fig. \ref{Fig:ExperimentalFitness}). None of the relative fitness values for the assayed haplotypes were statistically different from the AWT or WT.

The lowest relative fitness corresponds to the ancestral haplotype Q153E that is the consensus of the extant species. Although this difference is not statistically significant, it is a result that merits a comment. It has been argued \citep{Holmes2002,Carlson2014} that a consensus sequence of a protein in a group of species will often have a larger fitness since the fixed positions will be the result of selection for optimal configurations of the protein. Our result is in contrast to this finding. We return to this issue in the discussion.

\begin{figure}[p]
\centering
\includegraphics[scale=0.6]{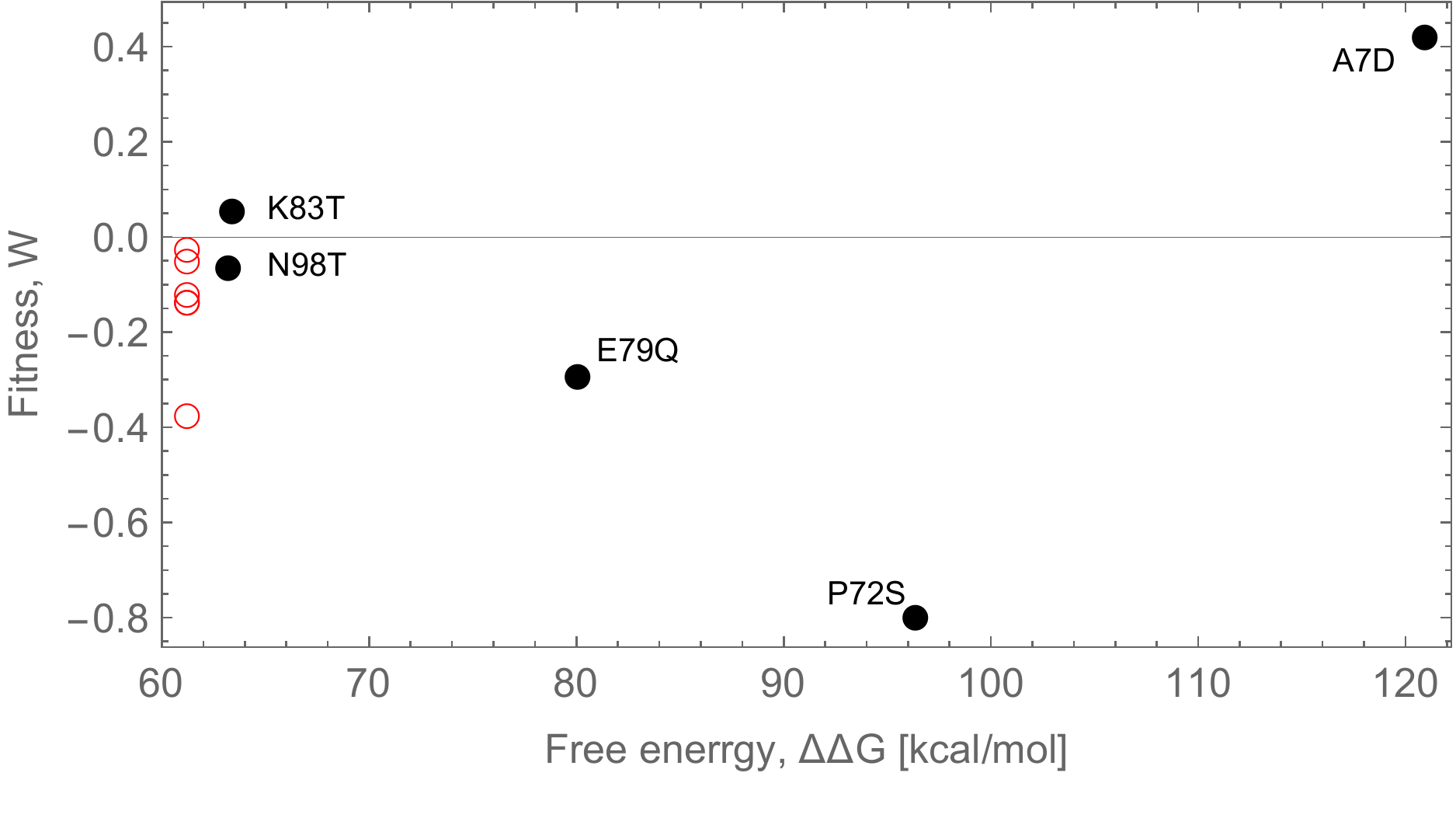}
\caption{Experimental fitness measures of \citet{Vale2012} against free energy. Bullets (black): amino acid substitutions; rings (red, color online): silent DNA substitutions.}
\label{Fig:FitnessValeEtal}
\end{figure}

\section{Discussion}
\subsection{Thermal stability vs steric compatibility}
FoldX has previously been used in the design of thermostable proteins \citep{VanderSloot2006} in the prediction of mutational effects to binding energies \citep{Kiel:2005fv}, and to estimate structural epistasis, in a similar way as we do in this work \citep{Bershtein2006}. It has been observed that the consensus sequence of a group of extant haplotypes shows higher thermal stability \citep{Imanaka1986,Lehmann2000a}, a reason why phylogenetic methods have become popular amongst structural biologists. 
 
However, we believe that thermal stability does not play any role in the evolution of the capsid of \phix . If thermal stability would be a relevant factor in the evolution of the capsid we would expect a pervasive decrease of free energy along the phylogeny. In fact, we found no evidence for any selective preference for substitutions that decrease the free energy.

Although free energy is typically associated with structural stability this interpretation requires some clarification. To be able to assess stability we would require free energy measurements not only of the actual configuration of the capsid but also in an alternative state as, for example, when the elements of the capsid are unassembled. Additionally, even given these measurements, we would still lack information regarding the activation energy that is required to disrupt protein-protein interactions and thus disassemble the capsid. This is because while free energy differences dictate the preferred direction of the conformational change, the activation energy dictates the expected waiting time for this change to happen.

In contrast the free energy FoldX measures refers only to the structural degrees of freedom of the side chains. A likely consequence of the free energy changes resulting from AA substitutions is that they can induce significant conformational changes in the capsid protein even before the assembly of the pro-capsid. If some of these conformational changes are large enough the pro-capsid cannot be assembled.

\subsection{Distribution of mutational effects}
Quantitative trait loci (QTLs) and genome wide association analyses (GWAS) have revealed that additive effects can be explained by right skewed distributions \citep{Visscher:2012je,Hindorff:2009cc} such as an exponential or gamma. Also there are compelling theoretical arguments justifying these empirical observations \citep{Keightley:2007hq,EyreWalker:2007dl}. Despite the biophysical nature of our trait, the mutation effects on free energy have a skew normal distribution, which is consistent with the shape of traditional quantitative models.

	One argument that explains said skewed distributions is that substitutions of large effect are strongly selected because they bring the trait closer to an optimum. Once close to an optimum, substitutions of small effects are favored over large ones because they allow fine-tuning of the trait to match the optimum trait value. Consequently there are few opportunities for mutations of large effects to fix and many more for the ones with small effects. This argument does not require that the effects are mediated biologically and clearly applies to a physical source of mutational variability, as in our case.

	The distribution of the effects in the random mutant set approximates that of a neutral process. This distribution is wider (with larger mean and variance) than that of the effects along the phylogeny (Fig. \ref{Fig:FreeEnergyDist}). By comparing both distributions, we can readily rule out that the latter is neutral \citep{Orr:2009wv} (Kolmogorov-Smirnov test $p = 10^{-7}$, Appendix \ref{SI:Selection}). This supports the hypothesis that selection acts on random mutational effects.

\subsection{Selection}
A similar argument to that of thermal stability has been proposed for fitness. It has been noted and experimentally determined that the consensus sequence tends to have the higher fitness \citep{Holmes2002,Carlson2014}. This is because most substitutions tend to be detrimental, and, under relatively low mutation rates derived alleles are not represented in the consensus sequence. However we point out that this might be the reason behind some reported exceptions to the consensus rule.
 
As argued throughout this article, structural information can help framing a picture about the causes of evolution of the \phix's capsid.
A first scenario is that the capsid is under stabilizing selection and substitutions that significantly deviate the structure from its optimum energetic value are purged out. In this case we expect the extant haplotypes to have similar energy distribution to the ancestralsÕ, which is what we observe. A second piece of evidence favoring this hypothesis is that the distribution of free energies observed in the random simulations, including multiple substitutions, is much wider than the one derived from ancestral reconstruction on the phylogeny. The molecular data, which shows a low average $dN/dS$ ratio ($\omega= 0.060-0.084$), indicates strong purifying selection, which is consistent with the stabilizing selection scenario.

An alternative scenario is that directional selection maintains the structure very close to an absolute energetic minimum. Unless there is substantial load, the occurrence of negative mutational effects is unlikely because no substitutions can further diminish the free energy of the capsid. Evidence against this scenario is that the class of mutations that diminish the free energy is not necessarily small. 

One could still assume directional selection with high mutation rates resulting in high load. In this case we can still expect mutations that can decrease the free energy to be observed. However high mutation rates would result in a larger spectrum of fixed substitutions that should result in high phylogenetic divergence. The low divergence observed in our data is indicative that selection is much stronger than mutation, which altogether argues against the directional selection scenario.

Indeed only 14 sites show a high posterior probability of having $\omega > 1$, while 315 sites belong to a category with an extremely low median for $\omega = 0.084$. Of the eight sites that were polymorphic in the ancestral state, six are under positive selection (there is a 7th site also with $\omega > 1$, albeit with low posterior probability). This corroborates the scenario where the coat protein is under very strong purifying selection, with very few available sites allowing for mutations to fix along the phylogeny.

As stated in the Results section (see also Fig. \ref{Fig:AminoAcidEffects}), histidine and leucine tend to have the strongest effects on \ddg. Thus, we expect that multiple substitutions to histidine would result in considerably large deviations of the free energy. We thus substituted all the ancestral polymorphisms (the four fixed and the eight polymorphisms at the ancestral node) to histidines, and compared it to a protein with 12 substitutions to histidines at random sites (S1H, G57H, F124H, E178H, A198H, G246H, M283H, F291H, G321H, G377H, Q405H, D421H). The former does not result in a free energy deviation that is nearly as bad as the latter ($\Delta \Delta G=213$ kcal/mol vs $\Delta \Delta G=735$ kcal/mol, respectively). This difference is expected from the point of view of structural biology and is also consistent with our hypothesis that purifying (stabilizing) selection is acting on the capsid.

A subtle point deserves some attention. If sites are under positive selection they will show $\omega > 1$. However, the converse is not always true. Instead, if sites are free of functional constraints and become unrestrained to change, they can have $\omega > 1$, even if these are not under positive selection \citep{Zhang2005}. This later scenario seems to be the relevant one for our case. That is, even when most of the ancestral polymorphic sites have $\omega > 1$ our experiments show that the AWT has no statistical difference in fitness against WT. This scenario also explains why the \ddg \, with the 12 ancestral polymorphisms mutated to histidines is much smaller than that of the protein with 12 randomly placed histidines. 

Further evidence for this effect comes from investigations on the distribution of mutational fitness effects of \phix \, \cite[][see also Fig. \ref{Fig:FitnessValeEtal}]{Vale2012}. More than two thirds of 36 random single point mutations (including synonymous mutations) resulted in fitness changes. Of all substitutions, only 5 non-synonymous mutations occurred in the coat protein (gene F). Of these five, two (K83T and N98T) have no significant effect on fitness, yet both have $\omega > 1$ in our phylogeny. The other three changes (A7D, P72S and E79Q) have $\omega < 1$ and show a significant effect in fitness, with A7D being a beneficial mutation. This pattern is further reflected in a significant negative correlation between fitness effects and $\omega$ ($R^2 = -0.84$).

The data on free energy also supports the hypothesis that selection has acted on the capsid. First, we argue against a neutral mutation accumulation scenario. On the basis of random mutations with a distribution of effects skewed to the left (as it is the case; Fig. \ref{Fig:FreeEnergyDist}A) we expect an increase of the average energy along the phylogeny, for which there is no supporting evidence. In contrast, under purifying selection, such as stabilizing fitness, the energy distribution stays centered at an optimal value, irrespective of the skewness of the mutational effects on the free energy. This is supported by the fact that the average of the free energy has not significantly increased along the phylogeny. However, its variance has; this is expected on the basis of genetic drift, although we may not discard that factors such as unequal mutation rates and/or indirect selection on correlated traits can be partly responsible for the extant variability. 

In the Appendix \ref{SI:Selection} we present a rough estimation of the fitness landscape equating it to gaussian stabilizing selection. For this matters we only employ the distribution of ancestral haplotypes and the distribution of random substitutions  (Fig. \ref{Fig:FreeEnergyDist}A, inset). We infer that the selective strength acting on an haplotype is about $S\simeq 0.005$, and an optimum centered close to $\Delta \Delta G \simeq -10$ kcal/mol. This optimum value coincides with the peak of the extant species. We stress that we did not include the extant species in the estimations; the coincidence in the peak values strongly supports the hypothesis of stabilizing selection. 

By ranking all the ancestral haplotypes according to how close their free energies are to this optimal value we identify which ones are the fitter. The closest is (K83Q, T92S, E150Q, Q182L, S339A) with \ddg $=-10.6$ kcal/mol. However, this has a very low probability ($1.2\times10^{-3}$) of being the true ancestor. The consensus sequence is in the 6th closest haplotype to the peak (\ddg $=-9.8$ kcal/mol), and the most likely haplotype is the 54th closest to the peak (\ddg $=-13.4$ kcal/mol). There is not any obvious correlation between this ranking and the probability of the ancestral haplotype, or with its probability rank.

If stabilizing selection is indeed acting on the coat protein of \phix \, capsid what could be its source? As argued before, steric constraints prior to pro-capsid assembly might be more limiting than stability of the capsid. Equally important is the ability of a protein to be able to undergo particular conformational changes that are important biologically. On the one hand, if substitutions introduce significant increase of the free energy, there might be steric constraints due to conformational changes. If the free energy is decreased too much, the protein, besides being sterically impaired, will also be more rigid. The interplay between these two steric factors is a possible source of stabilizing selection.

\subsection{Epistasis}
Epistasis facilitates mutation accumulation by compensating energetic deviations. In the Results section we report the existence of a correlation between epistasis and the number of mutations in evolved lineages. This correlation is stronger in the extant species (slope= 2.69, corr=0.35) compared to the rest of the data (slope=0.48, corr=0.016), reflecting an evolutionary increase of epistasis in the \phix \, family. 

It is widely thought, largely on the basis of theoretical population genetics arguments, that selection is more likely to favor antagonistic (or negative) epistasis over synergistic (positive) epistasis (Rice 1988; \citep{Desai:2007bq}. Under the assumption that most mutations have deleterious effects on fitness, epistatic effects that compensate the fitness decrease have higher fixation probabilities and, if under stabilizing selection, result in a lower genetic load. Our results indicate exactly the opposite trend: we find a positive correlation between free energy and epistasis. This seems to contradict with the hypothesis of stabilizing selection because epistasis enhances the capsidÕs energy rather than compensating it.

The antagonistic epistasis hypothesis assumes that there is a genotype that can achieve the optimum. In some models, it is also assumed that the distribution of epistatic effects can bring a trait arbitrarily close to the optimum. It might be that none of these two assumptions hold in our case. If an optimal energy cannot be attained due to physical constraints and there is a negative deviation from the optimum, positive epistasis can be favored \citep{Hermisson:2003dw}. This can explain the positive correlation between energy and epistasis. An alternative explanation for positive epistasis would be disruptive selection, but under this scenario we would expect a bimodal distribution of energetic effects, which is not the case.

\subsection{Physical basis of epistatic effects}
The distribution of epistatic effects in the ancestral haplotypes is centered nearly at zero and has a small variance compared to the distribution of free energy. This is in contrast to the extant species, whose distribution shows a larger variance (Fig. \ref{Fig:EpistasisDistribution}). Nevertheless, in general, epistasis contributes to the energy only by a small fraction of the additive values. What determines this distribution?

Because most haplotypes have only a few substitutions, the derived amino acids end up being physically distant. This results in weak electrostatic interactions because several components of the force fields decrease rapidly with the distance, and consequently the resulting epistatic effects are small. Conversely, larger number of mutations implies larger chances for any two substitutions to occur spatially close. The closer two amino acids are, the stronger their energetic interaction is, and therefore show strong epistasis. Also, as more mutations accumulate, we expect more interactions to be established because there is a larger combinatorial space, which allows higher order epistasis to act. Furthermore, spatially asymmetric forces, such magnetic dipoles and quadrupoles, can come into play, adding to the non-additive component of the free energy even more. This physical explanation can account for the increased correlation of the epistatic effects observed in the extant species.

This structurally based epistasis is somewhat different than the source of epistasis found in quantitative genetic systems, such as bacteria and higher organisms, because in the present case the effect of a substitution is entirely determined by physical laws. However, it remains true that the energetic contribution of any given allele (amino acid at a certain position) is sensitive to the background on which it appears, principally because of the interactions with its surrounding, but also due to other energetic terms (which we have not explicitly considered here but are included in the free energy calculations) such as torsional potentials, entropic contributions and also environmental effects such as temperature, ionic forces, pressure and other physical-chemical quantities. Therefore, the dependency on the background amino acid composition and on the environmental norm is entirely analogous (even if a somewhat reductionist picture) to that of quantitative traits.

Our notion of structural epistasis follows directly from a physical mapping from sequence space to free energy. This differs from other measures of epistasis based on statistical regressions. Typically, it is methodologically impossible, due to limitations of experimental design, to gather enough data to measure beyond the second or third epistatic order. Unlike our calculations of structural epistasis, statistical epistasis estimations can result in biases do to several methodological issues (e.g. \citet{Otwinowski:2014hk}). This is certainly an advantage because our approach does not make any approximations due to truncation of higher order epistatic coefficients.

\subsection{High order epistasis can compensate pairwise interactions}
We have estimated the epistatic components of the variable substitutions at the ancestral state up to fifth order. Most prior work has only considered pairwise epistasis; the usual assumption is that higher order terms are of lower effect or even negligible. However, the distributions of 2nd to 5th order epistasis can be of comparable magnitude, revealing a strong non-additivity. In all cases the epistatic coefficients are normally distributed and, roughly speaking, ranging between -1.50 and 1.50. kcal/mol (Appendix \ref{SI:Bonferroni}). Consequently, analyses based on only 2nd order epistasis can be severely biased \citep{Weinreich:2013iy}.

Statistical epistasis also supports the occurrence and relevance of high order epistasis. However, it is interesting that a predictor of the free energy that considers only additive interactions is almost as a good as a predictor that includes high order (up to 5-way) regression coefficients. Curiously, this does not imply that epistatic interactions are absent; many of the regression coefficients (even those of high order) are statistically significant. Similarly, t-test on structural epistasis indicates the presence of non-additive effects. The interpretation that we give is that high order structural epistasis compensates pairwise interactions in such a way that, for some haplotypes, the free energy appears to be largely additive.

The question is thus whether this epistatic masking has any relevance. The answer depends on the context of the question. For predicting the energetic values of a given amino acid sequence, it makes little difference. However, epistasis is known to mask genetic variation, in what is known as cryptic genetic variance. This is important in the evolutionary context because it confers evolvability to the capsid. Cryptic genetic variance occurs when a given allele damps down the detrimental effect of other substitutions. Consequently, although there might be certain amount of heterozygosity in the population, there is no genetic variation on the trait. Although in our analyses we donÕt consider populations, there is evidence for cryptic epistasis, which might have been an important mechanism in the diversification of the \phix \, family.

\section{Concluding remarks}
It is well known that a sequence determines a proteinÕs structure through the interactions amongst the composing AAs (i.e., the protein folding problem). Although the existence of interactions is not surprising from this structural point of view, what cannot be foreseen are the kinds of interactions that should be prevailing in evolved macromolecular structures.
From the side of evolutionary biology, it is not trivial to rationalize the expected strength of additive and epistatic factors. Moreover, how these evolve is also an active research question that lacks a definite answer.
By considering both frameworks together, that is evolutionary biology and structural biology, we have determined the distribution of the epistatic factors that have been preferred in the evolution of the capsid of the \phix \, family.
Contrary to expectations based on physical reasoning, there was no significant average change of the free energy during the evolution of the capsid. However, we found an increase of structural epistasis which is better explained in terms of the evolutionary history and the biology of the virus, rather than on thermodynamical arguments. This might be surprising for physicists and unsurprising for biologists. In either case, the central message is that the joint analysis has allowed understanding the mode of evolution in a way that would have not been possible by approaches based on only one point of view of the problem.
Also, by employing structural simulations, we have overcome limitations that otherwise would impede estimating the full extent of epistasis. Our finding that high order epistatic factors can have effects as large as those of second order epistasis, indicates that there is no hierarchy on the order of epistasis. Nevertheless, the interplay amongst the different epistatic terms serves as a buffering factor to mutation accumulation. This results, at least on average, in the compensation of effects as indicated by a negligible gain in the mean free energy distribution. It is surely a big challenge to understand higher order epistasis from the perspective from both experimental and theoretical population genetics, but structural biology can further aid in this understanding.

It remains unclear what the relationship is between folding free energy of the capsid, function and fitness. However, the fact that we find a strong signal of selection on free energy provides  compelling evidence of such  interconnection, irrespective of how complicated its nature.

\section{Materials and Methods}

\subsection{Estimation of the phylogeny}
We retrieved from GenBank 22 sequences from members of the Microviridae family of ssDNA virus. 18 sequences of \phix \, {\it sensu stricto} and 4 outgroups -- G4, NC13, WA13 and $\phi$K (Fig. \ref{Fig:PhylogeneticTree}). The following criteria were used in selecting sequences for our analyses: 1) the sequences originated from wild isolates of the virus, and 2) they had complete genome sequences available. We also included the canonical \phix \, Sinsheimer/Sanger strain (J02482, dubbed the wild type: WT). We explicitly excluded sequences that were derived from experimental manipulations (e.g., experimental evolution) or whose origin was unclear.

	The major coat protein gene (geneF) was extracted from the sequences and aligned using ClustalW, implemented in the Alignment Explorer function of MEGA 5.2 \citep{Tamura2011}. The resulting alignments were used to reconstruct the phylogenies in MrBayes 3.2.2 \citep{Ronquist2012}.

	Nucleotide, amino acid and codon models were used for the reconstructions assuming a GTR+$\gamma$ site substitution model ($+\omega$ in the case of codon model) with flat Dirichlet priors for the substitution rates of the GTR model and unconstrained branch lengths. All other parameters were left at their default values \citep{Ronquist2012}. The search was carried on using 2 independent runs of 4 Markov Chains (1 cold and 3 heated) for $2\times10^6$ cycles, sampling every 200 cycles with a burn-in of 25\% of the samples for topology and parameters estimates (Codon model estimates ran for $3\times10^6$ cycles, sampling each 300th). All three phylogenetic reconstructions gave compatible trees with similar topologies (Fig. \ref{Fig:PhylogeneticTree}, Appendix \ref{SI:Phylogenetics}).

	We used PAML 4.8 \citep{Yang:2007aa} to estimate $dN/dS$ ratio ($\omega$) for each site in the alignment of gene F using a number of models for codon evolution (M0, M2, M7 and M8), on the consensus tree that resulted from MrBayes analyses described above. The $\omega$ value is expected to be 1 if there is no selection acting in that codon (neutral site), it is expected to be $< 1$ if the site is under purifying selection and expected to be $ > 1$ for sites under diversifying (positive) selection. The best-fit model for our data is model M8, which uses a discrete $\beta$ distribution (k=10 classes) to model classes with $0 < \omega < 1$ and one additional class with $\omega >1$. PAML uses the Bayes empirical Bayes approach (BEB) to calculate the posterior probability for sites belonging to each class.

\subsection{Ancestral state reconstruction}
The ancestral reconstruction for the last internal node (Node A, Fig. \ref{Fig:PhylogeneticTree}) was carried out in MrBayes using the topology of the trees estimated above for each data set (AA, nucleotide and codons). Since the codon tree was the best resolved tree (i.e., had the least number of polytomies, Appendix \ref{SI:Phylogenetics}), for the internal nodes we used the codon tree as a guide to specify constraints defining the nodes and reporting posterior probabilities for the ancestral states for each node. An independent run of MrBayes was performed for each node. 

We adopted a non-orthodox approach for the sequence reconstructions reported in this work, whereby we considered not only the most probable sequence for each node but, taking advantage of the power of Bayesian inference in handling uncertainty of the posterior probabilities, we generated a set of all possible ancestors for each node. A site was considered ÒfixedÓ in the ancestor node if the posterior probability of a state was $P>0.99$. And was considered polymorphic otherwise. Moreover we only considered polymorphisms that were in agreement among the reconstructions, and discarded those that were not predicted by the three methods.

\subsection{Molecular model of the capsid}
The atomic structure of the bacteriophage \phix \, capsid, used as a starting point for our model, was previously solved by X-ray crystallography at 3 \AA\,resolution \citep{McKenna1992}. The virion capsid has a T=1 icosahedral symmetry, i.e., it consists of a repeat of 60 identical asymmetric units (Fig. \ref{Fig:CapsidModel}). Each asymmetric subunit consists of three proteins: the major coat protein (protein F, 426 AA), the major spike protein (protein G, 175 AA) and the DNA binding protein (also termed minor spike protein; protein J, 37 AA) (Fig \ref{Fig:CapsidModel}C). Additionally, it contains a short (5 nt) DNA fragment. This DNA fragment was not included in the structural models.

To study the changes of the free energy of unfolding of the different haplotypes, we modeled a capsid fragment that consists of 12 proteins subunits: one focal capsid protein surrounded by 11 other capsid subunits,  as well as one minor and one major spike proteins (Fig. \ref{Fig:CapsidModel}). This complex represents 1/5 of the virion capsid. By modeling this complex, we take into account the influence of neighboring protein chains that might affect energy through protein-protein interactions. When considering substitutions to the capsid protein, all copies of the fragment were mutated accordingly.

Prior to the usage of the capsid fragment to compute free energy deviations by introducing substitutions, the complex of 12 proteins was optimized by minimizing its energy. We employed the Amber ff99SB-ILDN force field \citep{Lindorff-Larsen2010} using the GROMACS 4.5 molecular dynamics simulation package \citep{Pronk2013}. The energy minimization was first executed in vacuum followed by a minimization  using solvent explicitly. This energy-minimized structure was used as a reference and as a starting point for further energetic analyses in the whole capsid fragment (i.e., the AWT)

\subsection{Energetic analysis}
To evaluate the changes in the free energy of the different capsid haplotypes we employ the protein design package FoldX (v3.0 beta 5.1). FoldX estimates the free energy of a given structure using a semi-empirical method relative to a reference structure. It employs a force field that considers a weighted combination of physical and statistical energy terms that were calibrated to fit experimental \ddg \, values from mutational experiments \citep{Guerois2002a,Tokuriki2007,Tokuriki2008,Schymkowitz2005}. FoldX corrects its predictions by using the empirical rectification formula \[\Delta \Delta G_\text{exp}=(\Delta \Delta G_\text{FoldX}+0.078)/1.14\] \citep{Tokuriki2007}.

As a reference structure for our calculations we employed the capsid fragment described above, but included the four substitutions that are fixed at the ancestral node (I3V, H216R, L242F and V318A; see Table \ref{Table:Polymorphisms}). Since these substitutions also appear in all of the extant species except in the WT \phix , we considered that this haplotype (the AWT) is an appropriate reference structure. Consequently epistasis is also measured relative to this structure.

Since the calculations are for a fragment, we divided the output \ddg \, of FoldX by 12 to account for the energy of a single copy of the coat protein.

\subsection{Structural Simulations}
All structural simulations were carried at 25\degree C. Using pilot simulations we concluded that 15 or more replicates of each run are sufficient for the sample variance of \ddg \, to be stable. We performed variance ratio tests comparing bootstrapped distributions of \ddg \, with 5 and $n=10,15,20$ replicates. With as few as 10 replicates the variance test showed $p<10^{-22}$ in all three cases, and comparing the distributions with 15 and 20 replicates we found no significant difference (data not shown). Although this indicates that 15 replicates are enough, for most haplotypes we employed 20 replicates, thus increasing statistical power. For the single nucleotide substitutions we employed as many as 50 replicates.

We computed a sample distribution of free energies for each haplotype at each internal and ancestral node, of the extant species and of each single substitution occurring in the alignment. The latter is required to estimate the epistatic effects. However, in many cases, the single mutants also occur at the internal or ancestral nodes.

\subsection{Estimation of structural epistasis}
In general, epistasis is defined as non-additive effects on a trait. The trait of interest in this case is the free energy change \ddg \, of the capsid. The identity of capsid, in turn is entirely determined by its AA sequence, that is equivalent to the haplotype $\mathcal{H}$. Since we are able to computationally estimate the free energy contributed by each single substitution $i$ composing the $ \mathcal{H}$, we can estimate epistasis, $\epsilon$, as
\begin{equation}
\epsilon = \Delta \Delta G_\mathcal{H} - \sum_{i \in \mathcal{H}} \Delta \Delta G_i ~.
\end{equation}

However, the data from each haplotype and their corresponding single substitutions are not paired. Moreover, different simulations can have different number of replicates. Therefore we can only make an average estimate of the epistatic value $\bar{\epsilon}$. However, we can nevertheless test statistically whether epistasis is negative, positive or zero. For this purpose, we performed a T-test using the following statistic
\begin{equation}
t= \bar{\epsilon} / \sqrt{\frac{V_\mathcal{H}}{n_\mathcal{H}}+ \sum_{i \in \mathcal{H}} \frac{V_i}{n_i} }\end{equation}
where \(V_k = \text{var}(\Delta \Delta G_k)
\), is the sample variance of the free energies of the haplotype $k$ and $n_k$ is the number of replicates. We employed WelchÕs approximation for the degrees of freedom for unequal sample sizes.

\subsection{Estimation of statistical epistasis}
Statistical epistasis is estimated by formulating a similar model to that above, albeit employing epistatic terms explicitly. We express the free energy differences as
\begin{multline}
\Delta \Delta G= \sum_{i=1}^n \alpha_i X_i + 
\sum_{\substack{ i,j=1 \\  i\neq j}}^n \epsilon_{ij}X_i X_j  \\
+ \sum_{\substack{i,j,k=1\\ i\neq j\neq k}}^n\epsilon_{ij}X_i X_j X_k +\ldots+\text{error}
\end{multline}
where $\alpha_i$ are the additive factors, $\epsilon_{i\ldots}$ the epistatic factors and $X_i$ the incidence variables. We set $X_i=0$ for the AWT allele and $X_i=1$ for the derived allele. The factors $\alpha$ and $\epsilon$ are estimated using an ANOVA analysis. Unlike on the model of structural epistasis were we only use the average \ddg \, of each haplotype, for the design matrix of the ANOVA we employ all data points of the simulations (4476 data values).

We employed a maximum likelihood estimation of the regression parameters (implemented by the function LinearModelFit[] of \emph{Mathematica} 10.0.1.0). The design matrix of the regression depends on the order of epistasis. The data limits us to considering epistasis terms up to order 5. Thus estimated models with only additive effects (no epistasis) and epistatic models with 2 to 5 interactions. We also computed AkaikeÕs Information Criterion to determine which model is preferred.

\subsection{Experimental methods }
In order to assess the fitness of the variant haplotypes found in the ancestral reconstructions we synthesized 10 of the 256 possible ancestor haplotypes.

The synthetic sequences of ancestral versions of the gene F were obtained from Epoch Life Sciences Inc., for each of the eight single variants (K83Q, T92S, P141A, E150Q, Q153E, Q182L, S339A, A361V), for the AWT and the AWT$^{(8)}$. These sequences were cloned into the double stranded form (RF1) of the wild type \phix \, (DSM4497, Sinsheimer/Sanger strain) replacing the wt gene F, and transformed into \emph{E. coli} C strain (DSM13127). 

To obtain the phage stocks the transformed \emph{E. coli} were mixed with non-transformed bacteria grown for 2 hours at 37\degree C to increase phage titers. The lysate was then centrifuged at 4\degree C (10,000g for 5 minutes) and filtered with a 0.22\micro M syringe-filter membrane to obtain the mutant phage stocks. These were then stored in glycerol at -80\degree C.

The titre of the \phix \, mutants stocks was determined using a soft agar overlay method. Early exponential growth phase ($\sim10^7$ cfu/\milli L) \emph{E. coli} C cells and L.B. soft agar (0.7\%) media, supplemented with 5mM of both CaCl$_2$ and MgCl$_2$, were mixed with serial dilutions, plated on LB agar (1.5\%) plates, and incubated overnight at 37\degree C. Plaque counting for each sample (mutants and the WT) were performed in 10 replicates with 2 plates/dilution. 

We used the growth rate as a proxy for estimating the absolute fitness of the phages as a measure of doublings of the population per hour in presence of excess numbers of bacterial host \citep{Bull2000}. The absolute fitness was estimated as 
\begin{equation}
r_t = \log_2\left[\frac{C_{t=60}}{C_{t=0}} \right]~,
\end{equation}
where $C_t$ is the concentration of the phage at measurement time $t$. Then we estimated Relative Fitness against a reference (both to the AWT or the WT) by taking the ratio of the absolute fitness of a given haplotype $r_i$ against the absolute fitness of the reference, $r_o$ \citep{Bull2000}.

In order to reduce measurement variance we performed structured experiments essaying each mutant paired with both the AWT and of the WT. The data reported in Fig. \ref{Fig:ExperimentalFitness} are averages over relative fitness measurements of these paired experiments, \(W_i=\left\langle \frac{r_i}{r_o} \right\rangle\).

The experiments always included two replicates of the mutant being essayed plus both the WT and AWT. The phages were diluted to $\sim 10^3$  [pfu/\micro L] based on the stock titer estimate in two replicates. 100\micro L of the replicates were mixed with 100\micro L of early-exponential growth phase \emph{E. coli} C (M.O.I = 0.0001) in 3\milli L of LB media. 

To obtain initial titers at $t_0$, for each replicate one aliquot of 100\micro L was immediately plated along with two step-dilutions from a second 100\micro L aliquot (see titration above). The remaining mixture was incubated at 37\degree C for 1 hour following purification (as for phage stocks), an aliquot of 100\micro L was taken to obtain the $t_{60}$ titer, diluted ($3\times 10$-fold steps) and plated (2 replicates per dilution). 

Each paired experiment was performed twice on different days. The experiments were performed using a double blind design, only revealing the identity of the assays once the experiments were completed. Each mutant had a total of between 16-24 replicates depending on the number of countable plaques on the plates (WT and AWT had 32-48 replicates), this experimental design accounts for variance in dilution, plating and handling during the experiments.

\section{Acknowledgements}
We would like to thank Profs. Marek Cieplak and Bojan Zagrovic for discussions and comments on the manuscript. HPdV was funded by a European Research Council grant ERC-2009-AdG for project 250152 SELECTIONINFORMATION. JPB is funded by a European Research Council grant ERC-2015-CoG for project number 648440 EVOLHGT.

%\bibliography{../phiX_paper_refs,../PhiX174refHPdV,../BookRefs,../PhiX_tomek}{}%%%refs.bib

\appendix
\section*{Appendices}
\section{Phylogenetics and Ancestral Reconstruction. \label{SI:Phylogenetics}}

Phylogenetic trees for the   $\phi \chi174$ coat protein were inferred using MrBayes using three data frameworks: amino acids (AA), nucleotide and codon datasets. The codon and nucleotide bayesian consensus trees are compatible and almost identical in topology with the nucleotide tree. Besides  two additional polytomies, they are otherwise identical clades. The AA tree has a larger number of unresolved branches and one incompatible clade with the codon and nucleotide trees.

However the ancestral reconstruction for the ingroup sequences (Node A in Fig. 1) has the same 8 ancestral polymorphisms and 4 fixed ancestral states as in the AA reconstructions. The codon reconstruction has an extra polymorphism at R101G in Node A  (although with very low probability) and the A361V substitution is missing. The nucleotide tree has an extra polymorphism in D338H.

Because the codon tree presented the best resolution, we chose it to specify the constraints for the reconstruction of the sequences at the internal nodes.

\begin{figure}
\begin{centering}
\includegraphics[width=\columnwidth]{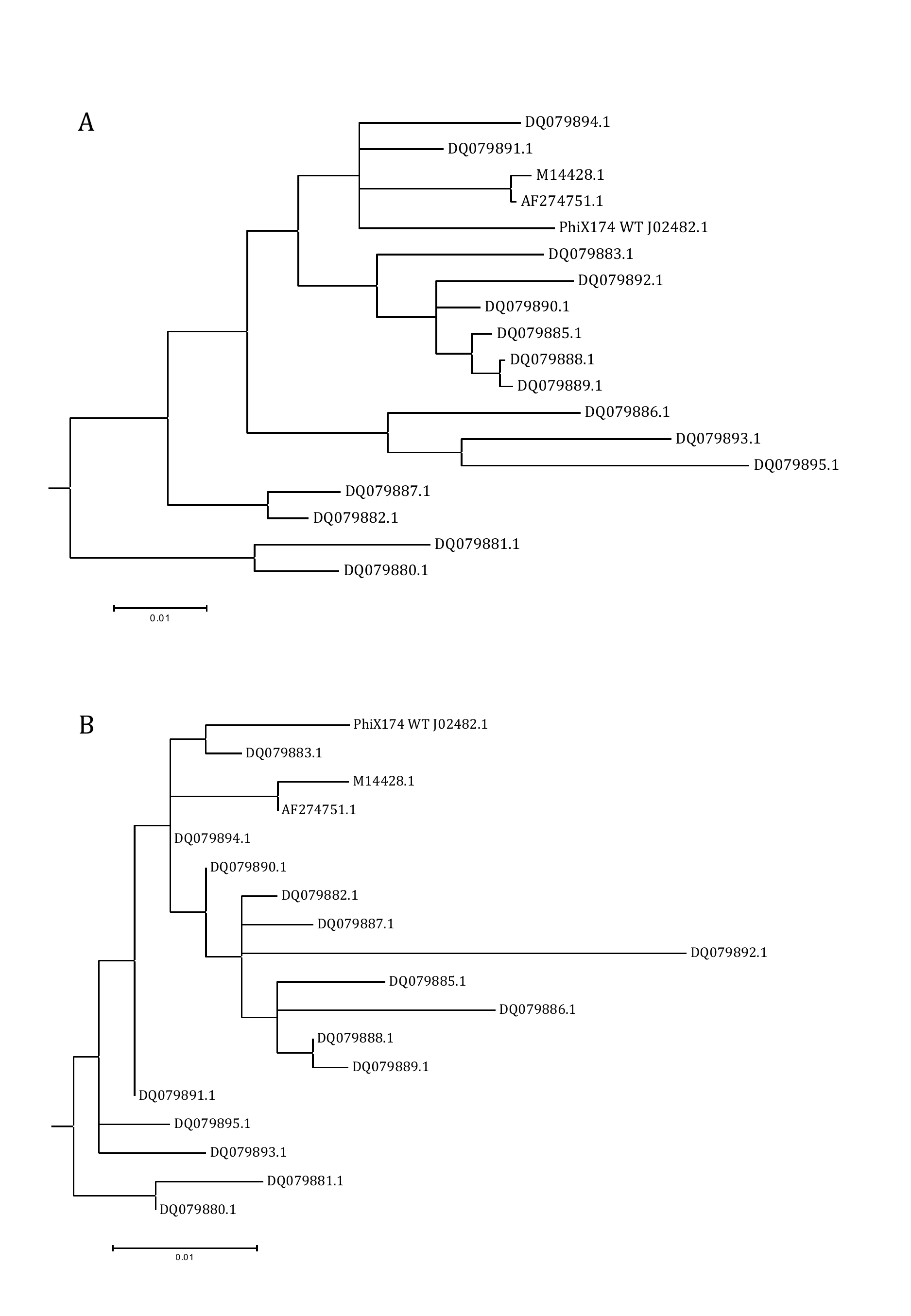}
\caption{Consensus phylogenetic of the  $\phi \chi174$ coat protein using MrBayes for: (A) Nucleotide and (B) amino acid datasets. The trees show only the ingroup sequences, although in the calculations the same outgroups as in Fig. 1 of the main text were used.}
\end{centering}
\label{}
\end{figure}

\section{Significance test of structural epistasis. \label{SI:Bonferroni}}

In this supplementary information we present details on our calculations of the T-test employed to detect epistatic haplotypes. First, we consider
that the average over samples of the free energy of a given haplotype $\mathcal{H}$ is

\begin{equation}
\overline{\text{$\Delta \Delta $G}}_{\mathcal{H}} = \sum _{k\in \mathcal{H}} \overline{\text{$\Delta \Delta $G}}_k+\epsilon _{\mathcal{H}}
\end{equation}
where \(\overline{\text{$\Delta \Delta $G}}_k\) are the means of the additive values (single mutants), and \(\overline{\text{$\Delta \Delta $G}}_{\mathcal{H}}\)
the mean of the multiple mutant. Thus \(\epsilon _{\mathcal{H}}\) is the average epistatic value.

The test is an extension of the standard T test for comparing samples. In this case we want to test the null hypothesis \(H_0\) that epistasis,
i.e.

\begin{equation}
\epsilon _{\mathcal{H}}= \overline{\text{$\Delta \Delta $G}}_{\mathcal{H}}-\sum _{k\in \mathcal{H}} \overline{\text{$\Delta \Delta $G}}_k ~ ,
\end{equation}
is significantly different than zero. For this purpose we define the standard deviation of the pooled sample 

\begin{equation}
S_{\mathcal{H}}=\sqrt{\frac{s_{\mathcal{H}}^2}{n_{\mathcal{H}}}+\sum _{k\in \mathcal{H}} \frac{s_k^2}{n_k}}~,
\end{equation}
where the \(s_k^2\) are the sample variances of \(\text{$\Delta \Delta $G}_k\).

In order to perform the t-test we also require the degrees of freedom \(\text{\textit{df}}_{\mathcal{H}}\). This is approximated with the Welch-Satterthwaite
formula, which reads

\begin{equation}
\text{\textit{df}}_{\mathcal{H}}=\frac{\left(\frac{s_{\mathcal{H}}^2}{n_{\mathcal{H}}}+\sum _{k\in \mathcal{H}} \frac{s_k^2}{n_k}\right)^2}{\frac{1}{n_{\mathcal{H}}-1}\left(\frac{s_{\mathcal{H}}^2}{n_{\mathcal{H}}}\right)^2+\sum
_{k\in \mathcal{H}} \frac{1}{n_k-1}\left(\frac{s_k^2}{n_k}\right)^2}~.
\end{equation}

\subsection{Bonferroni correction}

We need to apply the correction for multiple hypotheses tested on the same data. However, this number is not homogeneous across all tests because
there are independent data sets used.

Lets start with the simplest case. That is with the haplotype that has all indexes $\{$1,2,3,4,5,6,7,8$\}$. Since all indexes are involved, all
pairs, triplets, etc. are considered. We of course start count a 2 since there are no tests involving only one index. We have to add all unordered
pairs, all unordered triplets, and so on. This comes down to the number of subsets minus the sets with individual elements (minus the empty set).
Namely, 
\[m_8=2^8-8-1=2^8-2^3-1=2^3\left(2^5-1\right)-1\]
For haplotypes fixing a set of $k$ indexes, we need to count the number of subsets that have at least one of the indices. Note that this is more simply
calculated as the number of subsets of the remaining indices. There are $n-k$ unused indices of which there are \(2^{n-k}\) subsets. Hence, the number
of subsets that have at least one used index are \(2^n-2^{n-k}\). From this we finally subtract the number of sets with only one element, of which
there are exactly $k$. Thus
\begin{equation}
m_k=2^n-2^{n-k}-k~.
\end{equation}
This clearly matches the calculation above for $n=k=8$. 

\begin{figure}[t]
\begin{centering}
\includegraphics[width=\columnwidth]{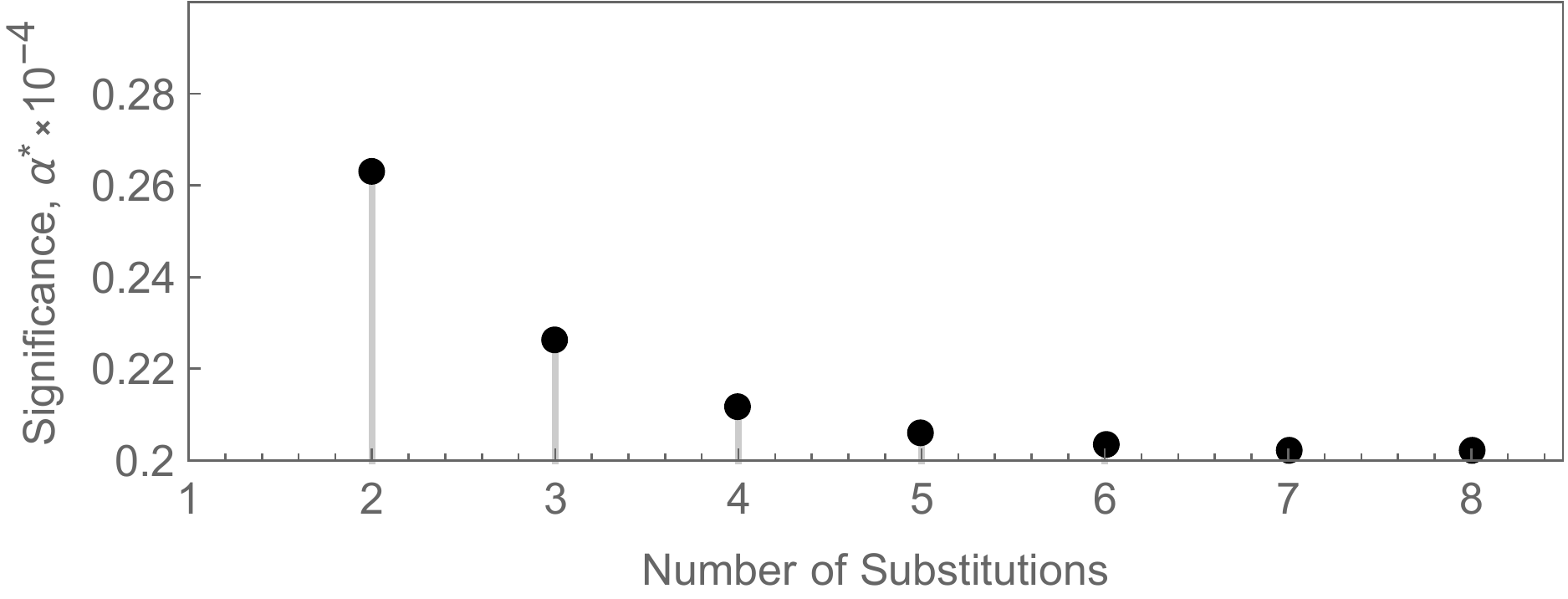}
\caption{Bonferroni correction for different haplotype numbers in our structured data set.}
\end{centering}
\label{Fig:BonferroniCorrection}
\end{figure}

The corrected significance for each test is therefore 
\begin{equation}
\alpha ^* = \alpha \left/\left(2^n-2^{n-k}-k\right)\right.
\end{equation}
where $\alpha $ is the base significance. Figure \ref{Fig:BonferroniCorrection}  shows the corrected significance for the current data and for a base significance of $\alpha
$=0.05.

All the above functions were tested and implemented in \textit{ Mathematica} 10.0.1.0.

\subsection{Results of the T-test}

\begin{figure}[t]
\includegraphics[width=\columnwidth]{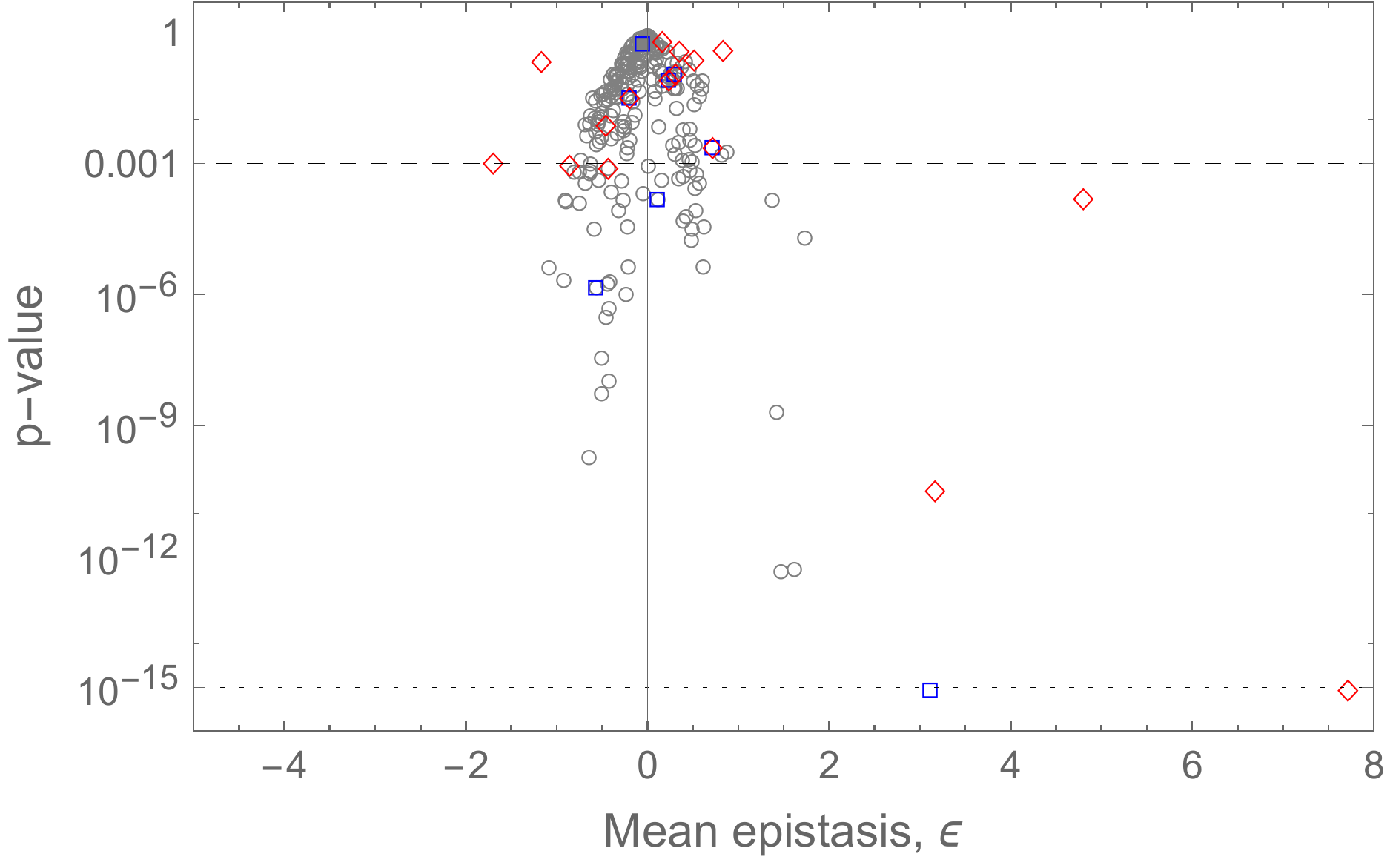}
\caption{Resulting p-values for our data of free energies $\Delta Delta G$ against the mean epistatic values  $epsilon$. Dashed line $p=1/256\simeq 0.002$. Dotted line: $p=10^{-15}$; $p-$values below this number are clipped in this figure. Grey rings: ancestral haplotypes; blue squares: internal nodes (other than ancestral); red diamonds: extant species.}
\label{Fig:ResultingPVals}
\end{figure}

Figure \ref{Fig:ResultingPVals} shows the results of the T test for our data set. 

\section{Inference of the selective landscape. \label{SI:Selection}}

Here we show by using theoretical and empirical arguments that the mutational variance generated by molecular substitutions is consistent with the stabilising selection scenario. Furthermore, this explains the distribution of free energies of the extant species. We also estimate the selective coefficient and the optimum value of the fitness landscape.

First of all, recall that selection is defined by the relationship
\begin{equation}
p_s(x) =  p(x) w(x)\,,
\end{equation}
where $p(x)$ is the frequency of the trait with value $x$ (e.g. $x\equiv\Delta \Delta G$), and the subscript $s$ indicated the distribution after selection; $w(x)$ is the relative fitness in the population of the trait value $x$. This implies that for every value of $x$, the fitness can be estimated by
\begin{equation}
w(x) = \frac{p_s(x)}{p(x)} ~.
\end{equation}

Our task is to have an estimate of $w(x)$, which in turn implies having an estimate of the distributions $p_s(x)$ and $p(x)$. If we asume that (a) the ancestral population has a similar distribution to the inferred distribution of ancestral haplotypes and (b) that this distribution is proximately the equilibrium distribution then we can take it as a sample for $p_s$. We take as $p$, the distribution before selection, as the distribution generated by mutational effects (see Results and Methods sections). However, for our calculations involving $p$ we only consider single substitutions. To facilitate our calculations we estimate these distributions using a gaussian kernel estimator. That is,
\begin{equation}
\hat{p}(x) = \frac{1}{n h} \sum_{i=1}^n \phi\left(\frac{x-x_i}{h}\right) \,,
\end{equation}
where $n$ is the number of histogram bins, $h$ is the `bandwidth', which accounts for the smoothing of the underlying histogram, and \(\phi(x)= \exp(-x^2/2)/\sqrt{2\pi}\) is the gaussian kernel.  The resulting smoothed distributions for the complete data (i.e. including multiple substitutions) are shown in the inset of the Fig. 5A in the main text with bandwidth $h=0.1$.

\begin{figure}[t]
\includegraphics[width=\columnwidth]{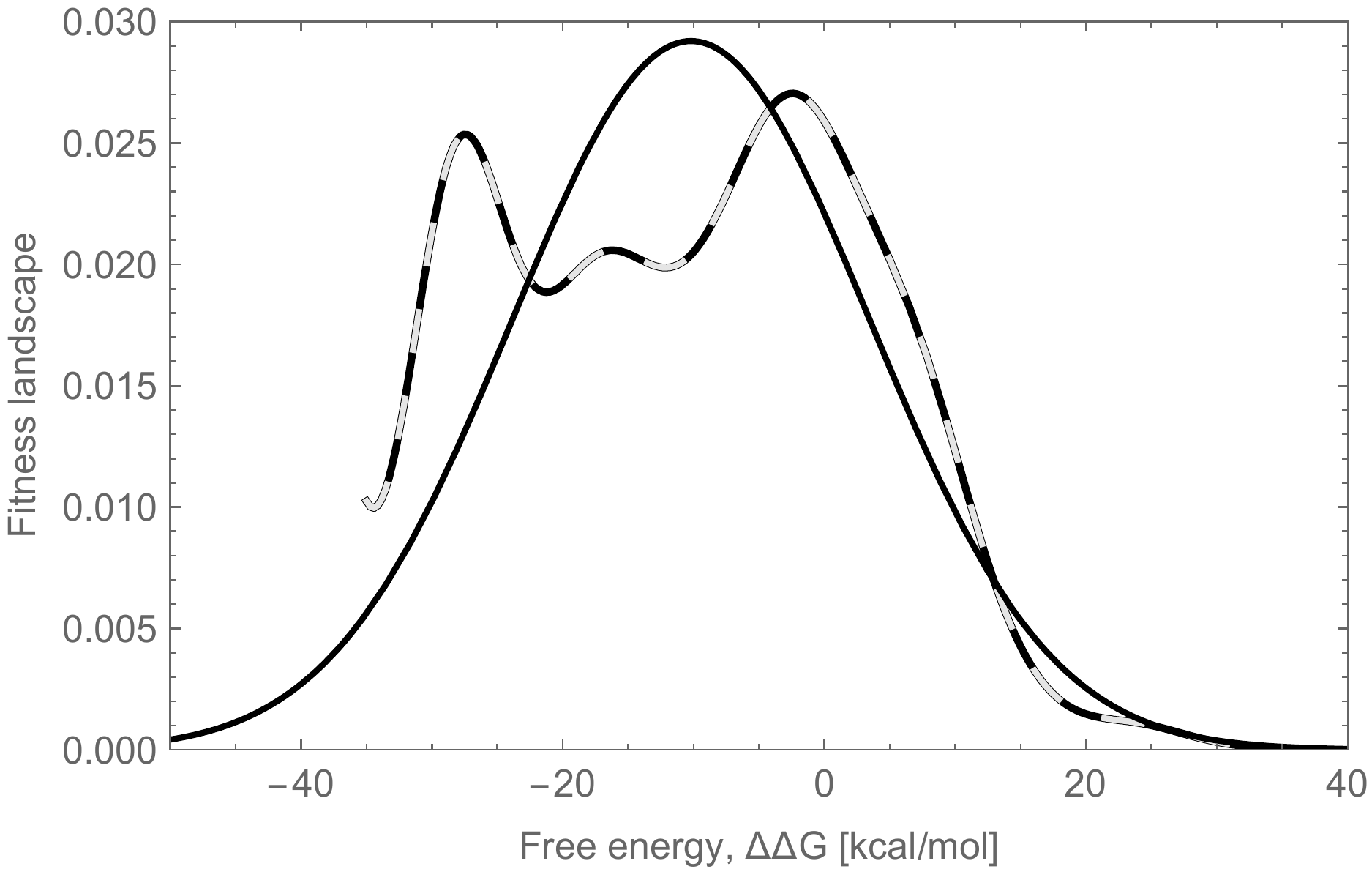}
\caption{Empirical approximation of a stabilising selection landscape. Dashed line: empirical landscape (using bandwidth $h=0.3$ and only data for single substitutions). Black: gaussian landscape with parameters inferred from the empirical landscape. Note that the scale of the landscape is of arbitrary units, and in this case both functions were normalised.}
\label{Fig:FitnessLandscape}
\end{figure}

Now se simply take the ratio \(r(x) = \hat{p}_s(x)/ \hat{p}(x) \), which is an empirical estimator of the fitness landscape $\hat{w}(x)$ (Fig. \ref{Fig:FitnessLandscape}, dashed lines).

We now assume that gaussian stabilising selection is acting, so that
\begin{equation}
w(x) = c \exp\left[ -\frac{S}{2}(x-x_{opt})^2 \right] \,.
\end{equation}
Here, $c$ is an arbitrary constant, $S$ is the selective value, and $x_{opt}$ is the optimal phenotype, in our case, the optimal free energy value. This parametrisation of the fitness landscape in terms of a gaussian function is standard in the literature of population genetics, largely for mathematical convenience. Although other forms are possible, the gaussian form captures the general relevant features of stabilising selection. Since we are interested only in rough estimates, we adhere to this standard for easiness.

Because the constant $c$ is arbitrary (since fitness is always with respect to the mean fitness of the population), we simply take a normalised version of $w$, so that $c=\sqrt{S/2\pi}$. This calls also to normalise the ratio $r(x)$, which we perform as a numerical integration. Because $w(x)$ has a gaussian form, the integral of its first and second central moments (mean and variance respectively) are equated to the first and central moments of $r(x)$. Using $h=0.3$, the mean gives us the optimum trait value \(\Delta \Delta G_{opt} \simeq -10\) kcal/mol, and the inverse of the variance gives the selective values $S\simeq 0.0054$. It is remarkable that this simple estimation of the optimum trait value coincides almost perfectly with the peak of the empirical distribution of free energies of the extant species (Fig. 5A in the main text), even though this data was not used in the estimation.

Figure \ref{Fig:FitnessLandscape} shows the empirical ratio and the inferred stabilising selection landscape. Clearly, this hasty calculation results in considerable deviations, obvious in the graphic. Yet, the gaussian landscape does reproduce well the general features, allowing for some robust inference. 

As a final note, the ratio $r(x)$ is sensitive to the bandwidth $h$. However there is little with the variation with $h$, where \(\Delta \Delta G_{opt} \simeq -20\) kcal/mol and $S\simeq 0.007$ for $h=0.1$. However, we favour the former estimate on the basis that it also matches the central mass of the distribution of free energies of the extant species.

\end{document}